\documentclass[]{aastex701}
\begin{document}
\title{La Silla-QUEST RR Lyrae Survey III: the shape of the Outer Virgo Overdensity}
\author{R. Zinn}
\affiliation{Department of Astronomy, Yale University}
\email[show]{robert.zinn@yale.edu}
\author{P.S. Coppi}
\affiliation{Department of Astronomy, Yale University}
\affiliation{Department of Physics, Yale University}
\email[show]{paolo.coppi@yale.edu}
\author{C. Baltay}
\affiliation{Department of Physics, Yale University}
\email[show]{charles.baltay@yale.edu}
\author{D. Rabinowitz}
\affiliation{Department of Physics, Yale University}
\email[show]{david.rabinowitz@yale.edu}
\author{Earl P. Bellinger}
\email[show]{earl.bellinger@yale.edu}
\affiliation{Department of Astronomy, Yale University}
\author{Kyra Bettwy}
\affiliation{Department of Astronomy, Yale University}
\email[show]{kyra.bettwy@yale.edu}
\begin{abstract}
Using the La Silla-QUEST and Gaia surveys, we identify 256 RR Lyrae stars (RRLS) in the Outer Virgo Overdensity (OVOD). These show that the OVOD has a signifcantly larger and more complex shape than previously recognized.  The OVOD lies at an average distance of $\sim 85$ kpc, has a long dimension of $\sim 57^\circ$ ($\sim 85$ kpc), and a width that varies from $\sim 10^\circ $ ($\sim 15$ kpc) to $\sim 35^\circ $ ($\sim 50$ kpc) in a prominent bulge.  Because it bends by $\sim 35^\circ$ from a great circle, it does not appear to be a stellar stream.  The type ab RRLS (RRab) span a large range in period, with a mean of $0 \fd 59$, and only $19\pm 3 \%$ lie near the locus of Oosterhoff II variables in the Bailey diagram.  The overall large number of RRLS along with a mean RRab metallicity ([Fe/H]) of -1.60 and $1 \sigma$ dispersion of 0.32, as inferred from the RRab periods, suggest that the OVOD progenitor was a large dwarf galaxy.  The hypothesis that the OVOD may be a shell produced by the Gaia-Sausage-Enceladus merger \citep{chandra2023} is supported by the $\sim 70^\circ$ angle that the southern extension of the OVOD makes with the orbits of red giants in its central part. The complex shape of the overall RRLS distribution, including a small north-south distance gradient, may also indicate, however, that we are seeing a blended, composite structure projected on the sky.     
\end{abstract}

\section{Introduction}
 Observations of the Galactic halo over the past 5 decades have provided increasingly strong evidence that most of the stars and globular clusters in the halo originated in dwarf galaxies that were disrupted while merging with the Milky Way (MW).  Similarly, numerous investigations over the last two decades have shown that the same is true of the halos of M31 and other galaxies. These results have become some of the strongest observational evidence in favor of the hierarchical picture of galaxy formation in which massive galaxies, such as the MW, have grown through a multitude of mergers with smaller systems.  The observational and theoretical studies that have led to this picture are discussed in reviews by, for example, \citet{majewski1993, freeman2002, mo2010, newberg2016, bullock2017, mcconnachie2018, simon2019, helmi2020, deason2024, bonaca2025}. 

Although the mechanism by which galaxies acquire their stellar halos has been arguably identified, important properties of the MW's merger history may be still unknown.  For example, there is continuing debate about what is widely considered the largest merger in the MW's history, the Gaia-Sausage-Enceladus (GSE) merger. It was firmly identified by \citet{belokurov2018} and \citet{helmi2018}, although some earlier evidence existed for a large, near-radial merger \citep[e.g.,][]{dinescu2002, brook2003, morrison2009, carlin2012}.  The GSE is thought to have happened $\gtrsim 8$ Gyr ago, because the stars in the GSE are of that age \citep[e.g.,][]{gallart2019, bonaca2020}, and because this merger is estimated to be sufficiently large to have puffed up a pre-existing thin disk into the thick disk, which is composed of old stars.  An alternative picture has been proposed by \citet{donlon2019, donlon2020}, who argued that the kinematics of the local halo stars are best explained if the merger, which they termed the Virgo Radial Merger (VRM), occurred $\lesssim 3$ Gyr ago.  On the basis of the chemical compositions of the stars, \citet{donlon2022} and \citet{liu2025} have argued that more than one large merger may have occurred. These issues and other pieces of the GSE ``puzzle'' have been reviewed in depth by \citet{deason2024}.  Clearly, much remains to be learned about it and other properties of the Galactic halo.

A large number of stellar streams and Galactic substructures have already been found \citep[e.g.,][]{bonaca2025}, but not many at large distances from the Sun ($\rm d_{\sun} \gtrsim 70$ kpc), where only a few surveys to date have had the sensitivity to probe large areas of sky. The two most significant of these are the Pisces Overdensity \citep[$\rm d_{\sun} \sim 85$ kpc,][]{sesar2007, watkins2009} and the Outer Virgo Overdensity (OVOD) \citep[$\rm d_{\sun} \sim 80$ kpc,][]{sesar2017}. Their origins have already been investigated in some detail \citep[e.g.,][]{carlin2012, chandra2023}, yet, as we demonstrate here,   
there is still more to be learned about them. More substructures also likely remain to be mapped in the outer halo.    

In this paper, we use the La Silla-QUEST (LSQ) RR Lyrae star (RRLS) survey to examine in more detail the OVOD. The discovery of the OVOD by \citet{sesar2017} consisted of finding 73 probable type ab RRLS (RRab) in the range $65 \leq \rm d_{\sun} \leq 95$ kpc that lie in an approximately elliptical region of the sky spanning $\sim 150 \,\rm deg^2$. Although the OVOD is close to the leading stream of the Sgr Stream when projected on the sky, \citet{sesar2017} argued that the OVOD is a separate halo substructure because it is $\sim 20$ kpc more distant than the Stream and offset from its mid-plane. \citet{sesar2017} used the term ``outer'' to distinguish this overdensity from the previously discovered and already well-researched Virgo Overdensity (VOD)\citep[e.g.,][]{vivas2001, newberg2002, juric2008, bonaca2012}.  The VOD is a much closer overdensity, $10 \lesssim \rm d_{\sun} \lesssim 25$ kpc, which may or may not be debris from the same merger event as the OVOD.  The VOD contains several moving groups of stars, and its properties motivated the VRM hypothesis by \citet{donlon2019}\citep[see also][]{carlin2012}.  Several authors \citep[e.g.,][]{simion2019, naidu2021, balbinot2021} have identified the VOD and the Hercules-Aquila Cloud (HAC) as debris from the GSE merger. Since the discovery of the OVOD by \citet{sesar2017}, it has been detected by surveys using blue horizontal branch stars \citep{amarante2024} and red giant stars \citep{viswanathan2026}, but they have not expanded its dimensions.  The LSQ RRLS survey shows that the OVOD is in fact a much larger overdensity than previously detected.

We start with a brief description of the LSQ survey, and then discuss the shape and dimensions of the OVOD, the properties of its RRLS, and what the new findings may mean for the possible origin of the OVOD.  In particular, we compare our results with those of \citet{chandra2023}, who have presented evidence that the OVOD is debris from the GSE. 
  
\section{The La Silla-QUEST RRLS Survey}

The LSQ variability survey was a 6 yr survey of the sky between declinations ($\delta$) $-80^\circ $ and $+25^\circ$ and galactic latitudes $|\rm b| \gtrsim 15^\circ$, with a primary purpose to search for supernovae and trans-Neptunian objects.  It was conducted on the 1 m Schmidt telescope of the European Southern Observatory at La Silla using the $10 \, \rm deg^2$ QUEST camera.  The camera and its broad-band filter are described in \citet{baltay2007} and \citet{zinn2014}.  Most of the LSQ exposure times were 60s, with a small minority of 180s ones.  The latest LSQ source detection, photometric reduction, and period-finding techniques, which are substantial improvements over those reported in \citet{zinn2014}, are described in detail in \citet{coppi2024}. They have enabled the measurement of the light curves of RRLS to magnitudes $\rm V \sim 21.5$, corresponding to distances $\gtrsim 140$ kpc, in regions of low interstellar extinction.  In Appendix A, we present a new calibration of absolute V magnitude ($\rm M_V$) as a function of logP, which is based on the RRLS in globular clusters with well-determined distance moduli and reddenings.  Using this calibration, the relative distances of the LSQ RRLS have a $1 \, \sigma$ error of $5\%$.

The completeness for RRLS and the LSQ photometric precision of the survey were described in \citet{coppi2024}, where comparisons were made between the survey and other studies of the RRLS in the Crater II and Sextans dSph galaxies, which lie at distances from the Sun ($\rm d_{\sun}$) of $\sim 116$ and $\sim 86$ kpc, respectively.  Our survey found 77\% of the known RRLS in Crater II and 93\% of the ones in the less distant Sextans.  It also discovered \citep[][]{coppi2024} the tidal tails of Crater II that span $80 \lesssim \rm d_{\Sun} \lesssim 140$ kpc, which indicate that tidal stripping has removed a significant fraction ($\gtrsim 30\%$) of Crater II's stellar mass.  Subsequent investigations have confirmed these results \citep[][]{limberg2025, vivas2026, atzberger2026}.  Comparisons with previous V band measurements of RRLS in Crater II revealed that our measurements of the intensity averaged V magnitudes, $\langle\rm V \rangle$, have at most a few hundredths of a magnitude offset from the standard Johnson V scale, and a standard deviation ($\sigma$) $\sim 0.05$ at $\langle \rm V \rangle \sim20.9$.  The fundamental mode (type ab or RRab) and the first overtone (type c or RRc) variables are identified and cataloged separately in our survey, but not the double-mode variables (type d or RRd), which are not easily recognized as a separate class without further analysis.  (The majority of the RRd's in our survey are likely classified as RRc, because the firstovertone mode is usually the one producing the largest amplitude.)         

The completeness of the LSQ RRLS survey depends in detail on the number of exposures obtained by the LSQ survey, which varies in a complicated way over the sky. In general, fewer observations were made near the upper limit in declination ($\delta = +25^\circ$) and the lower limit in galactic latitude ($|\rm b| \sim 15^\circ)$.  Even though the images were dithered between exposures, the gaps between the CCDs and the fact that only $\sim 85\%$ of the 112 CCDs in the camera were functioning meant that some objects within the same footprint of the camera were imaged fewer times than others. In addition, observations were made during conditions of variable seeing and transparency, which resulted in more acceptably good exposures for bright objects than for faint ones.

\begin{figure}
\plotone{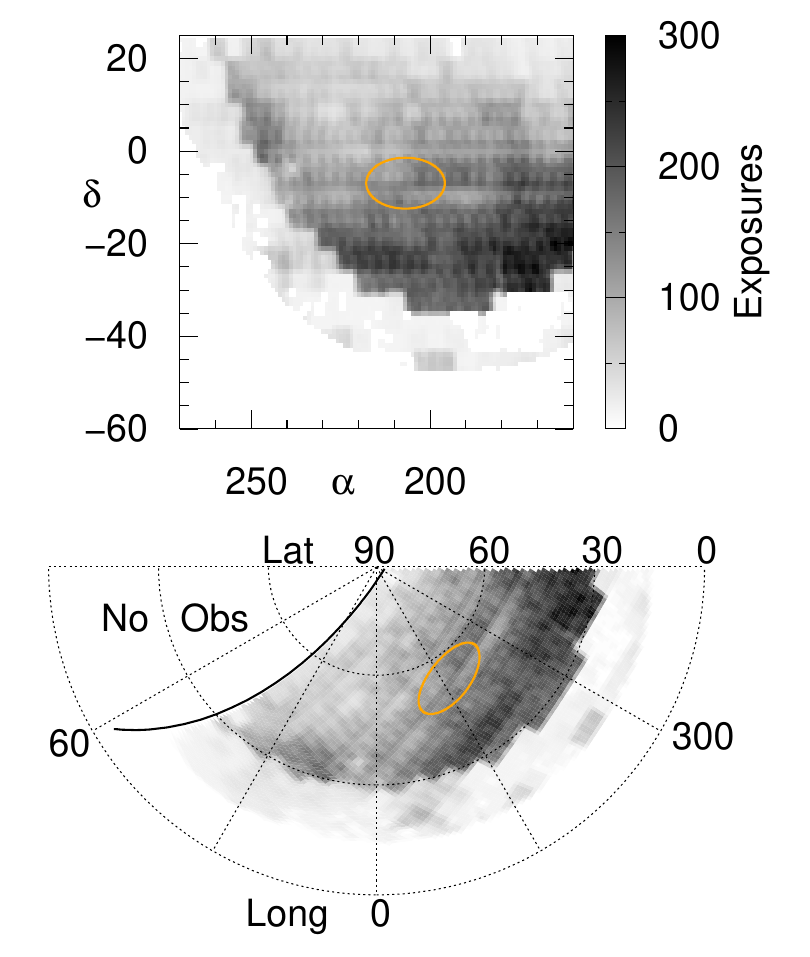}
\epsscale{1.4}
\caption{The number of good exposures for RRLS in the range $75 \leq \rm d_{\sun} \leq 100$ kpc is plotted as a gray scale in two projections of the sky.  Top: equatorial coordinates, bottom: galactic coordinates (azimuthal equidistant polar projection), both in degrees.  The orange ellipses mark the location of the OVOD region identified by \cite{sesar2017} (see Fig. \ref{fig1}). The LSQ survey did not cover the region labeled ``No Obs'' in the bottom diagram and in figures \ref{fig3}, \ref{fig4}, and \ref{fig6}. }
\label{figI}
\end{figure}

To illustrate how the number of LSQ exposures varied on the sky, in Figure \ref{figI} we plot two projections of the sky in the area around the OVOD.  The gray scale shows the average number of good exposures for the RRLS that lie in the range $75 \leq \rm d_{\sun} \leq 100$ kpc, the range that is most important for investigating the OVOD (see below).  These maps, which are averages over $1^\circ \,\rm x \, 1^\circ$ squares, take into account the airmasses of the observations and the interstellar extinctions \footnote{From \citet{schlegel1998} as revised by \citet{schlafly2011}.} at the sky positions.  They show that at the position of the OVOD, the number of good exposures was typically $\geq 100$.  In the expansion of the OVOD to the south (see below), the number was $\gtrsim 200$ until it plummeted at $\rm b \lesssim 30^\circ$. To the north of the OVOD, the number of good exposures was $\gtrsim 50$ until $\delta = +25^\circ$.  To find and determine the periods of RRLS in this magnitude range, a minimum of $\sim 30$ good exposures are required, which for $\rm b > 30^\circ$ was obtained nearly everywhere.

\begin{figure}
\plotone{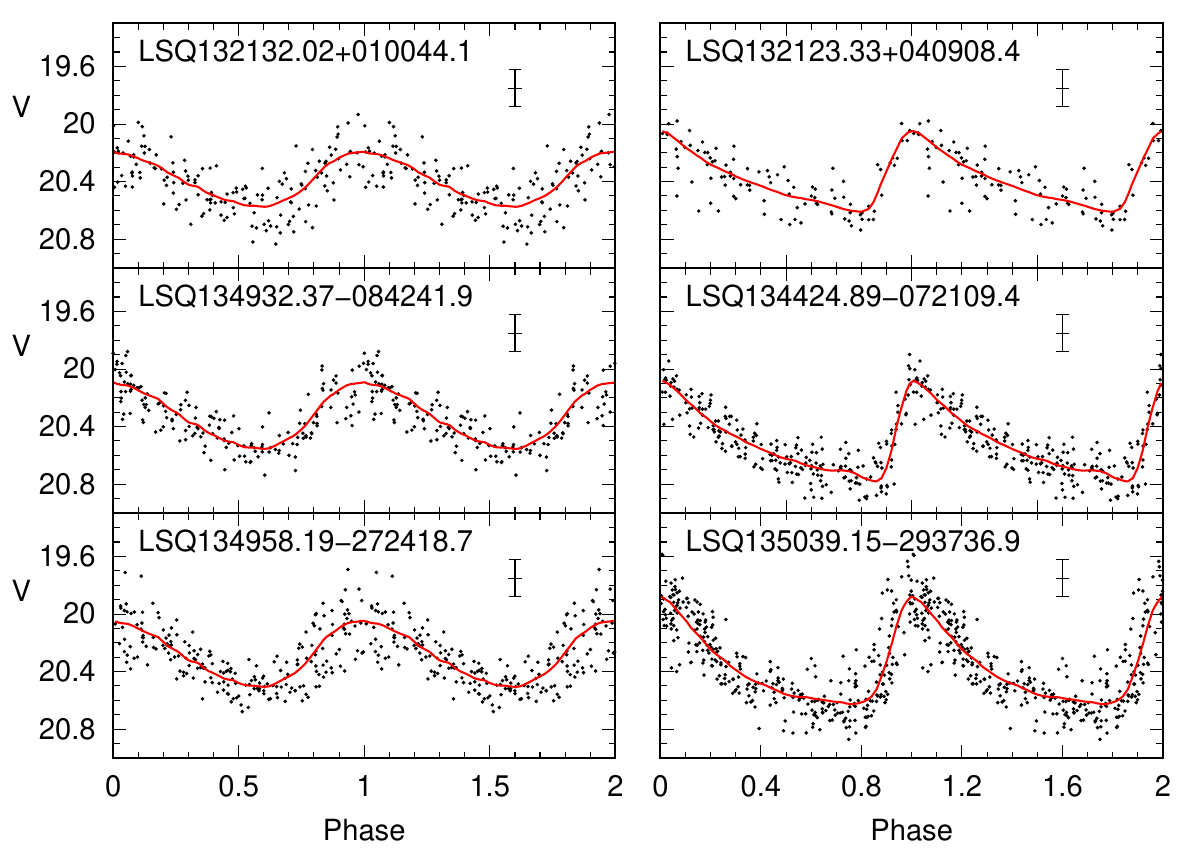}
\epsscale{1.8}
\caption{The phase-folded LSQ light curves of 6 RRLS in the OVOD are plotted along with their template fits (red curves).  The stars in the left and right columns are RRc and RRab variables, respectively.  The top, middle, and bottom rows are stars in the subdivisions of the OVOD: OVOD-N, OVOD-C, OVOD-S, respectively (see Section 3.2). The error-bars are $\pm 1 \sigma$ errors of the individual observations.}
\label{figlc}
\end{figure}

In Figure \ref{figlc}, we plot the light curves and template fits for 6 typical RRLS in the OVOD and its extensions beyond the orange ellipse plotted in Figure \ref{figI}, which are described below.  At the distance of OVOD ($\rm d_{\sun} \sim 85$ kpc), the individual LSQ observations have substantial errors ($\sigma \sim 0.1$). Because of the large number of observations, the phase-folded light curves are nonetheless well determined, as are the quantities measured from the light curves: period (P), amplitude ($\Delta \rm V$), and  $\langle \rm V \rangle$ ($\sigma \lesssim 0.001, 0.05, 0.05$, respectively).   See Fig. \ref{fig:gaia+lsq} for an example of how this works and a comparison with results from Gaia.    
     
\section{The RRLS in the Outer Virgo Overdensity}
\begin{figure}
\plotone{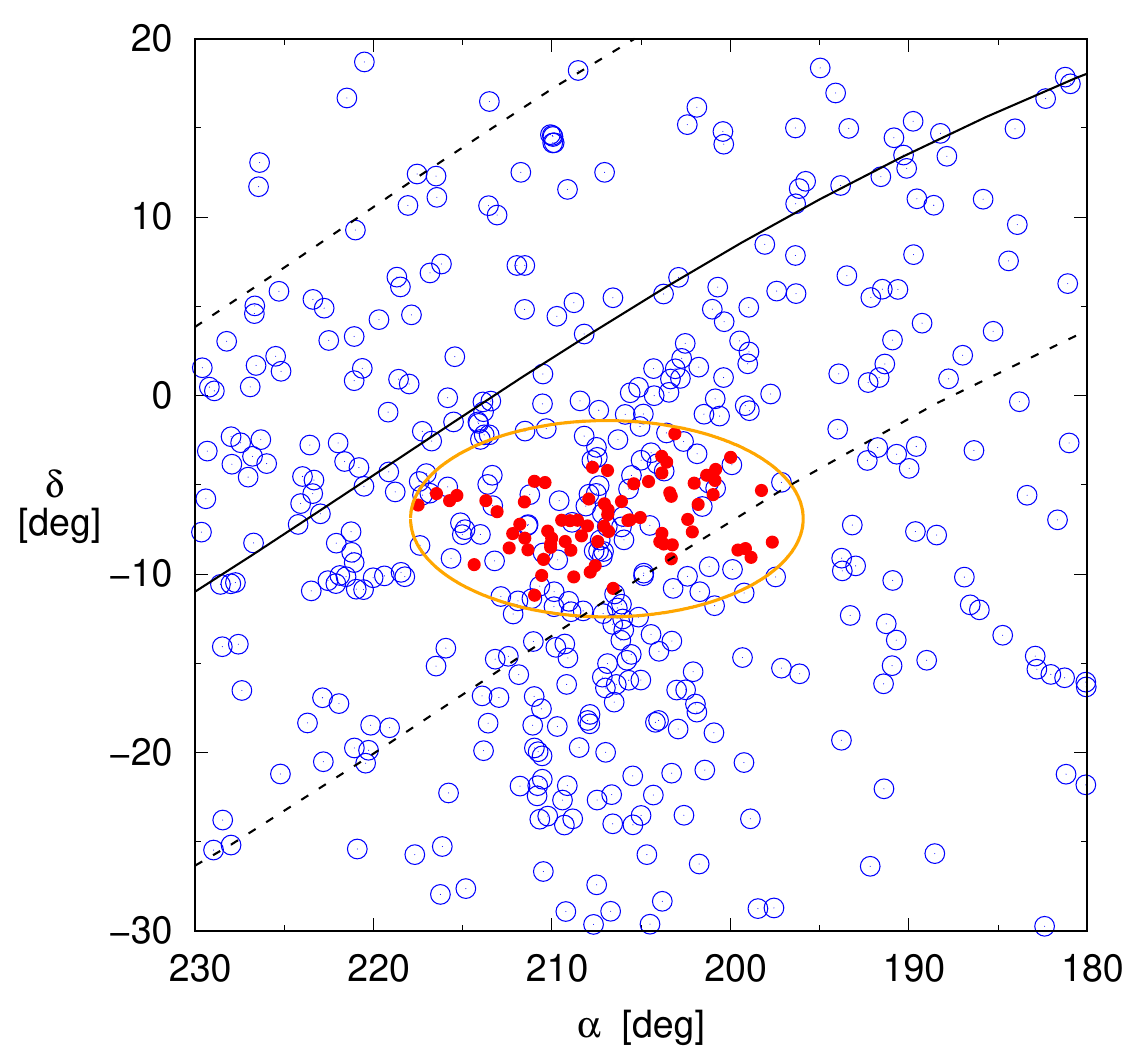}
\caption{The positions of RRLS in the sky surrounding the OVOD region identified by \cite{sesar2017}.  The \textit{red points} are the 73 RRab variables that were identified as OVOD members by \cite{sesar2017}, which lie at distances of $65\text{--}95$ kpc.  To account for the slightly longer distance scale used here, $65\text{--}100$ kpc was used to select the other RRLS in the region from the LSQ survey (\textit{open blue circles}). As used in other diagrams, the {\it orange} ellipse approximately delineates the region of sky containing the \cite{sesar2017} OVOD RRLS. The {\it solid} line shows the center of the Sgr Stream \citep{belokurov2014}, and the {\it dashed} lines show its approximate edges, $\pm 13^\circ$ from the center-line.  The roughly linear feature at ($\alpha,\delta$) = ($182^\circ,-15^\circ$) is part of the leading stream from the Crater II dSph galaxy \citep{coppi2024}.  The Bo\"otes I dSph galaxy is at approximately ($210.0^\circ, 14.5^\circ$).}
\label{fig1}
\end{figure}

The 73 probable RRab variables that led to the discovery of the OVOD by \citet{sesar2017} lie within the range $65 \leq \rm d_{\sun} \leq 95$ kpc and occupy an approximately elliptical region of the sky of $\sim 150 \,\rm deg^2$.  These 73 stars are the red points in Figure \ref{fig1}, where the orange ellipse is our depiction of the OVOD region, as identified by \citet{sesar2017}.  The LSQ RRLS survey recovered 70 of these stars.  For most of the stars in common, the periods are in excellent agreement, which is remarkable because \citet{sesar2017} had too few observations to make full light curves. According to the LSQ survey, two of the stars are more likely to be of type RRc than RRab as classified by \citet{sesar2017}.   For one of the variables missed by the LSQ survey, the number of LSQ observations (12) is too small to determine a period.  The other two missed stars are either not variable or very low amplitude according to the LSQ observations. For the stars in common, the LSQ distance values of $\rm d_{\sun}$ are on average $\sim 3$ kpc larger than those given by \citet{sesar2017}. To accommodate this offset, we thus extend the distance range of the LSQ RRLS positions plotted in Figure \ref{fig1} to $65 \leq \rm d_{\sun} \leq 100$ kpc.  

In addition to the 70 RRLS in common with \citet{sesar2017}, we see from Fig. \ref{fig1} that the LSQ survey discovered 76 more RRLS in the range $65\leq \rm d_{\sun} \leq 100$ kpc that also lie inside the orange ellipse (46 RRab and 30 RRc).  These stars and the other RRLS in the same range of $\rm d_{\sun}$ are plotted as open circles in Figure \ref{fig1}.  Also depicted in this figure is the location of the leading Sgr Stream, which lies in the foreground of the OVOD.  While the largest density of points is clearly in the OVOD, the LSQ survey reveals for the first time a region of high density that extends from the OVOD to more southern $\delta$ and outside the region of the Sgr Stream.  There is also evidence that the region of high density extends somewhat to the north and west of the OVOD. Since this is near the middle of the area occupied by the Sgr Stream, it is essential to see if it is part of the Stream.

\begin{figure}[h!]
\epsscale{1.2}
\plotone{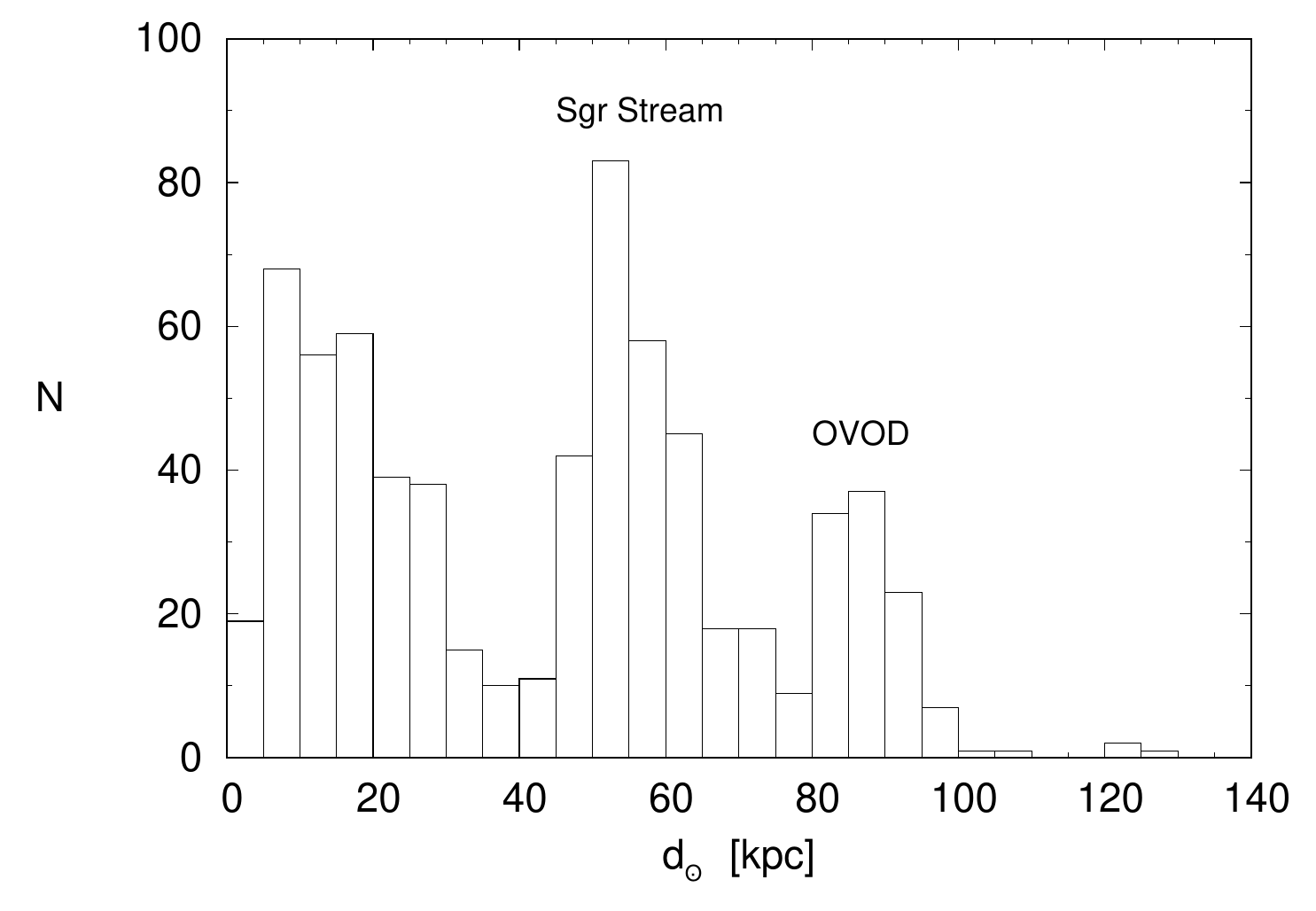}
\caption{The distribution in 5 kpc bins of the distances to the 694 LSQ RRLS that lie inside the OVOD sky region delimited by
the orange ellipse in Figure \ref{fig1}, which is approximately the location of the overdensity identified by \citet{sesar2017}.}
\label{fig2}
\end{figure}

\begin{figure}[h!]
\epsscale{1.0}
\plotone{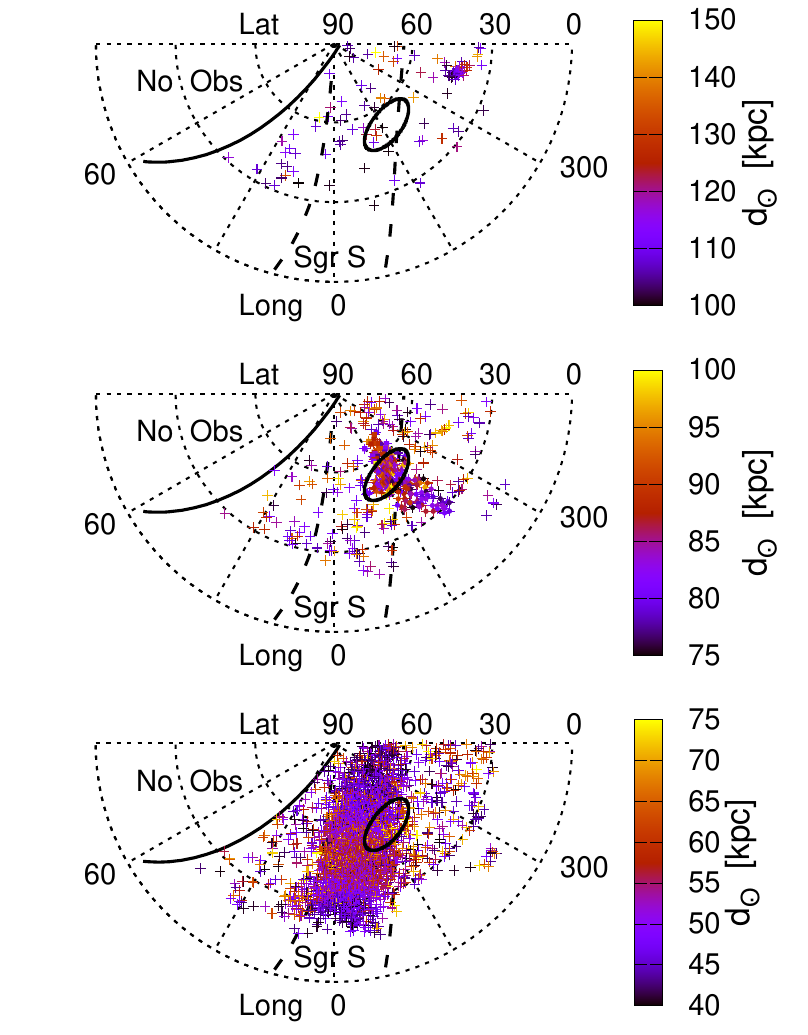}
\caption{For 3 ranges in $\rm d_{\sun}$, the positions of LSQ RRLS are shown in polar plots of their galactic coordinates.  The solid black line shows the limit of the LSQ observations ($\delta \leq +25^\circ$), and ``No Obs'' refers to the region not observed. The dashed lines are $\pm 13^\circ$ from the center of the Sgr Stream (``Sgr S'').  The bottom diagram shows that a vast majority of RRLS in $40 \leq \rm d_{\sun} \leq 75$ kpc lie between the dashed lines.  The middle diagram shows that the densest concentrations of stars are in the OVOD (\textit{black ellipse}) and both to its south and north.  The top diagram shows that for $\rm d_{\sun} > 100$ kpc there are few RRLS in the OVOD.  The Crater II dSph galaxy and its tidal tails can be seen at (l,b) $\sim 283^\circ, +42^\circ$ and $80 \lesssim \rm d_{\sun} \lesssim 140$ kpc.}
\label{fig3}
\end{figure}

In Figure \ref{fig2}, we plot a histogram in distance of all 694 LSQ RRLS inside the OVOD ellipse.  One readily sees that there are two significant peaks in the distribution.  The largest is due to the Sgr Stream, which begins at $\rm d_{\sun} \sim 40$ kpc.  It may extend to $\sim 70$ kpc, and to avoid its contamination, we consider the OVOD starting at $\rm d_{\sun} = 75$ kpc in Figure \ref{fig3}.     

For three ranges of RRLS distance from us ($\rm d_{\sun}$), we now plot in Figure \ref{fig3} the positions of LSQ RRLS in polar plots of galactic coordinates (l, b) using azimuthal equidistant projections.  Lines of longitude are therefore straight lines and all points are in the correct direction from the pole.  There is, however, some enlargement in this projection of the area of features as one approaches the galactic equator from the pole.  Since most of the LSQ observations are at $\rm b \gtrsim 25^\circ$, this limitation is tolerable. The bottom diagram illustrates the wide and dense Sgr Stream, which is by far the most significant overdensity in this part of the halo.  The color coding of the points reveals that the Sgr stream reaches its maximum $\rm d_{\sun}$ of $\sim 60$ kpc near the position of the OVOD. In the middle diagram, $75 \leq \rm d_{\sun} \leq 100$ kpc, the density in the Sgr Stream is greatly diminished, and the densest region is now the OVOD and its extensions to both more southern and northern latitudes. The previously detected OVOD, the black ellipse, is in fact only the densest part of a long feature that extends $\gtrsim 50^\circ$ in b.  The top diagram in Figure \ref{fig3} shows that few RRLS more distant than 100 kpc are likely members of this feature. 
\begin{figure}[h!]
\plotone{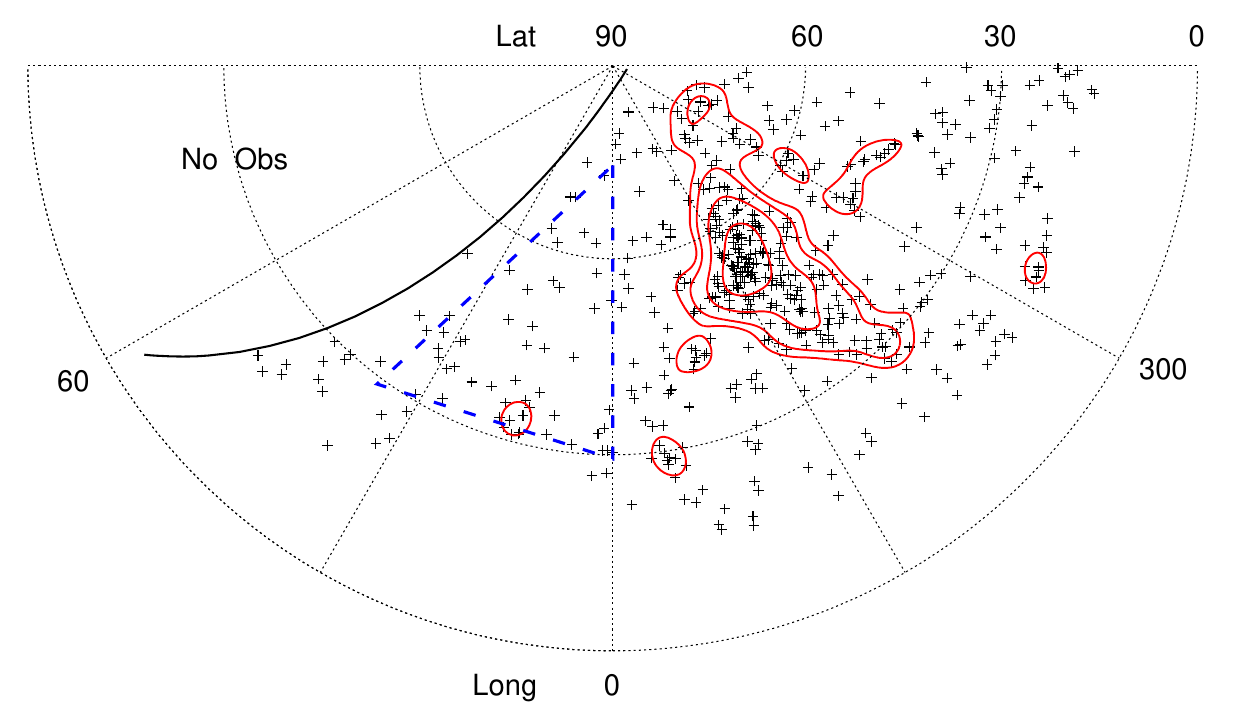}
\caption{This azimuthal equidistant projection shows in more detail the sky distribution of the RRLS from the LSQ and Gaia surveys in the range $75 \leq \rm d_{\sun} \leq 100$ kpc. The stars within the triangle (\textit{blue dashed lines}) were used to estimate the density of stars that are not associated with the OVOD.  The 4 contours (\textit{red curves}) enclose overdensities that are more than 3, 5, 10, and 20 $\sigma$ above this background density.  The 256 RRLS that we identify with the OVOD lie within the $3\,\sigma$ contour that surrounds the original detection of the OVOD.}
\label{fig4}
\end{figure}

Figure \ref{fig3} shows that the extension of the overdensity to the south of the black ellipse ends at $\rm b \lesssim 30^\circ$.  Because this might be an artifact of the precipitous drop in the completeness of the LSQ survey at $\rm b < 30^\circ$, we carefully selected RRLS from the Gaia catalog and placed them on the same distance scale as the LSQ stars (see Appendix B).  This procedure added 148 more RRLS with $75 \leq \rm d_{\sun} \leq 100$ kpc in the region of the sky depicted in Figure \ref{fig3}, bringing the total in that distance range to 564. Many of the Gaia stars not identified by LSQ do indeed lie at $\rm b \leq 30^\circ$, as anticipated from the poor LSQ coverage there. The Gaia stars are included in the position plot shown in Fig. \ref{fig4} (but not in Fig. \ref{fig3}, which only shows stars from LSQ). 

\subsection{The Size and Shape of the OVOD}

Figure \ref{fig4} is a plot of the LSQ plus Gaia RRLS in the range $75 \leq \rm d_{\sun} \leq 100$ kpc. The red overdensity contours in the plot were constructed from a Gaussian mixture model of the RRLS sky density, and to set the significance levels of the contours, we considered the density of RRLS within the blue triangle to be the ``background'' of RRLS in this distance range.  This region does not contain known halo substructures.  The contours in Figure \ref{fig4} correspond to levels of 3, 5, 10, and 20 $\sigma$ above the mean density within the blue triangle, where $\sigma$ is the standard deviation of the density. The OVOD region discovered by \citet{sesar2017} lies within the highest density contours. Note, however, that  the region of significant sky overdensity in fact extends an additional $\gtrsim  20^\circ$ to the north and south of this previous detection. A total of 256 stars are enclosed in the $3 \, \sigma $ contour (236 LSQ + 20 only from Gaia).  The locations and properties of these stars are given in Table \ref{tab:rrl_list}. These stars constitute $45 \%$ of the whole sample of RRLS in this region of the sky with $75 \leq \rm d_{\sun} \leq 100$ kpc.  These contours suggest that the OVOD does not extend to the south much beyond $\rm b \sim 27^\circ$. The $3\, \sigma$ contours also encircle several other separate and small overdensities. The one at $\rm l \sim 289^\circ , \rm b \sim 48^\circ$ includes the part of the leading stream from the Crater II dSph galaxy \citep[see][]{coppi2024}, which suggests that some of these smaller overdensities may also be real and not just noise.

Figure \ref{fig4} shows that the $3\,\sigma$ contour enclosing the OVOD starts at $\rm (l,b) \simeq (285^\circ,80^\circ)$, undergoes a $\sim 35^\circ$ bend at $\simeq (325^\circ,52^\circ)$, which is near the point of the highest density of RRLS, and ends at $\simeq(314^\circ,27^\circ)$.  The $5\,\sigma$ contour and to a lesser degree the $10\,\sigma$ contour follow the same bend, which strengthens the case that this is a real feature of the overall OVOD overdensity. The physical dimensions of the $>3\sigma$ overdensity are quite large.  The northern ``leg," i.e., $\rm b > 52^\circ,$ is $\sim 45$ kpc in length, and the southern ``leg'' $\sim 39$ kpc.  The width is $\lesssim 15$ kpc, except in the large bulge near the middle where it expands to $\sim 50$ kpc.  There is a small gradient in $\rm d_{\sun}$ with position along the OVOD in the sense that the northern ``leg'' is more distant than the southern one.  At the northern and southern extremes, the mean $\rm d_\sun$ is $\sim 87 $ and $\sim 82$ kpc, respectively.  

The bend in the contours in Figure \ref{fig4} is unexpected if the OVOD is simply a distant stellar stream, in which case it should follow, at least approximately, a great circle on the sky. A good example of stream morphology is provided by the Sgr Stream, which over the area of the sky covered by Figure \ref{fig4} nearly follows a straight line (see Fig. \ref{fig3}).

\subsection{The OVOD is not a spur of the Sgr Stream.}

\citet{sesar2017} argued that the OVOD is unlikely to be part of the Sgr Stream because it is separated in $\rm d_{\sun}$ and because the center of the OVOD is offset from the plane of the Sgr Stream by $\sim 9^\circ$.  The LSQ observations provide additional evidence that they are unrelated.

\begin{figure}[h!]
\epsscale{1.1}
\plotone{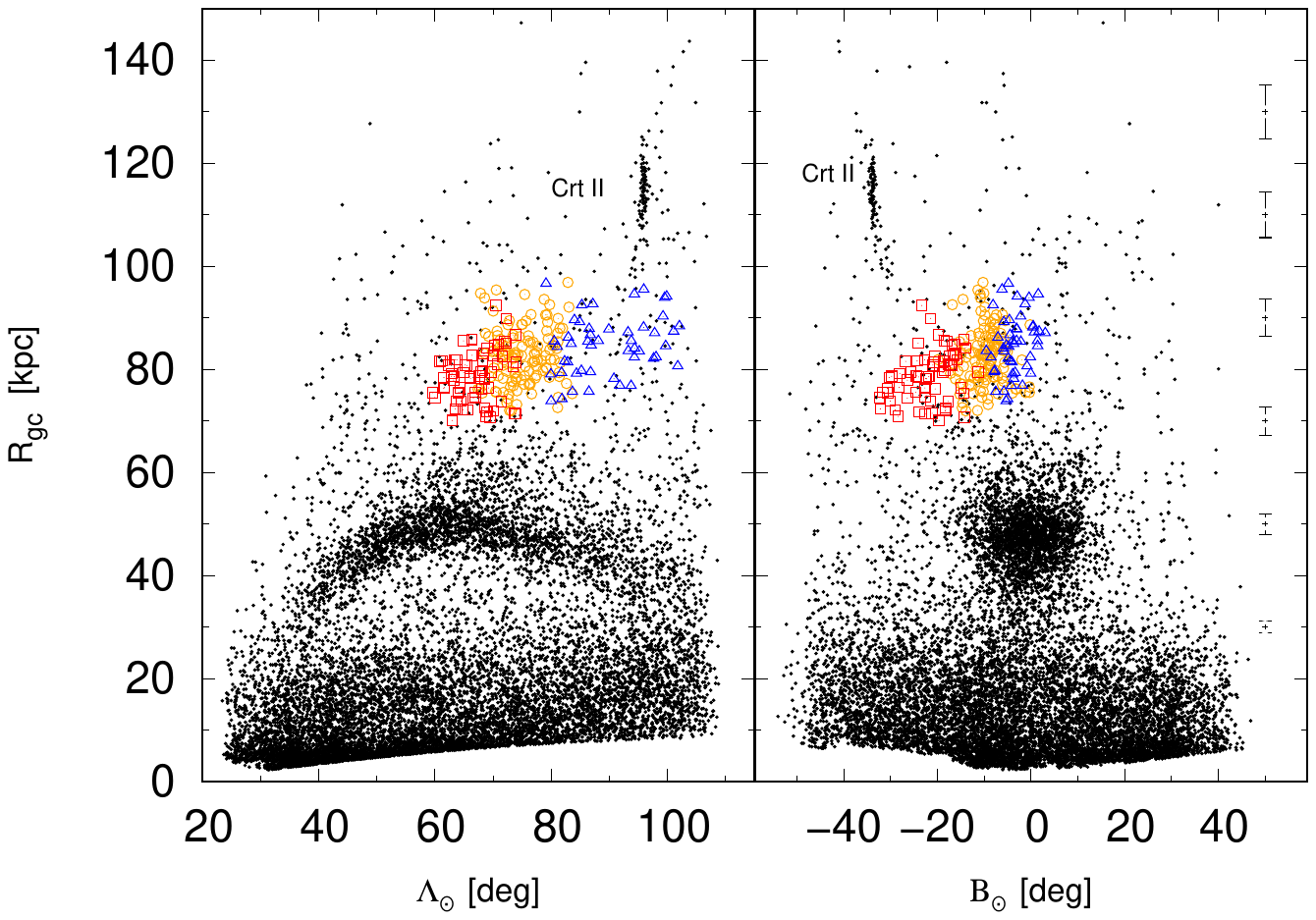}
\caption{For the LSQ RRLS within the same galactic longitude range as the previous diagrams (i.e. $l \geq 270^{\circ} \, \rm{and}\, l \lesssim 50^\circ$), galactocentric distance, $\rm R_{gc}$, is plotted against Sgr Stream longitude on the left and latitude on the right \citep{belokurov2014}.  The \textit{red open squares, the orange open circles, and the blue open triangles} are, respectively, the OVOD-S ($\rm b < 46.4^\circ$) , OVOD-C ($46.4^\circ<\rm b<59.1^\circ$), and OVOD-N ($\rm b > 59.1^\circ$) galactic latitude RRLS samples defined in the main text. The Crater II  dSph galaxy (Crt II) and its tidal tails are also evident in the plots.  The error bars are the $\pm 1\,\sigma$ errors in $\rm R_{gc}$.}
\label{fig5}
\end{figure}

Figure \ref{fig5} is a plot of galactocentric distance ($\rm R_{gc}$) against Sgr Stream longitude ($\Lambda_{\sun}$) and latitude ($\rm B_\sun$) \citep{belokurov2014}, using the whole sample of LSQ RRLS in the same region of the sky as Figures \ref{fig3} and \ref{fig4}.  The LSQ RRLS within the $3 \, \sigma$ contour in Figure \ref{fig4} with $ 75 \leq \rm d_\sun \leq 100$ kpc have been divided into three groups.  The central group (OVOD-C) is essentially the sample within the ellipse that was plotted in previous diagrams to mark the location of the original detection of the OVOD by \citet{sesar2017}.  It is the densest part of the OVOD.  The southern (OVOD-S) and northern (OVOD-N) groups consist of OVOD RRLS with $\rm b < 46.4^\circ$ and $\rm b > 59.1^\circ$, respectively.     

The plot on the right of Figure \ref{fig5} shows that the OVOD-N and OVOD-C samples overlap with the Sgr Stream in $\rm B_{\sun}$, but the OVOD-S does not.  The plot on the left shows that where there is overlap in $\rm B_{\sun}$, there is also one in $\Lambda_{\sun}$. But at these positions, there is also a wide separation in $\rm R_{gc}$, which in the case of the OVOD-N sample is $\gtrsim 30$ kpc. There is no sign of a bridge of stars between the Sgr Stream and any part of the OVOD. The Sgr Stream is concentrated within $|\rm B_\sun| \leq 13^\circ$, but the OVOD is clearly not, because it extends from $\rm B_\sun \sim 0^\circ$ to $\sim -30^\circ$.  Consequently, the OVOD deviates substantially from the plane containing the known structures of the Stream, including its distant spurs \citep[e.g.,][]{bayer2025, bayer2026}. Simulations of the Sgr Stream \citep[e.g.,][]{dierickx2017, fardal2019, vasiliev2021}, which have predicted or reproduced the distant spurs, do not predict one at the position and $\rm d_{\Sun}$ of the OVOD.   
\begingroup
\setlength{\tabcolsep}{3pt}
\begin{deluxetable*}{lrrcrrrrrrrl}
\tablecaption{Properties of Individual OVOD RR Lyrae Stars}\label{tab:rrl_list}
\tablehead{
\colhead{Variable ID} & \colhead{$\alpha$} & \colhead{$\delta$} & \colhead{Type} & \colhead{Period} & \colhead{$\Delta V^\tablenotemark{a}$} & \colhead{${\rm HJD}_{max}^\tablenotemark{b}$}  & \colhead{$\langle V \rangle$} & \colhead{$A_V$} & \colhead{$N_{fit}^\tablenotemark{c}$} & \colhead{$d_\sun$} & \colhead{${\rm Source}^\tablenotemark{d}$} \\[-0.2cm]
\colhead{} & \colhead{(deg)} & \colhead{(deg)} & \colhead{} & \colhead{(d)} & \colhead{(mag)}
& \colhead{(d)} & \colhead{(mag)} & \colhead{(mag)} & \colhead{} & \colhead{(kpc)} 
& \colhead{(survey)} 
}
\startdata
LSQ123129.67+132456.2 & 187.87364 & 13.41562 & RRab & 0.54610 & 0.98 & 56785.99495 & 20.18 & 0.07 & 93 & 80.7 & LSQ \\
LSQ123256.37+144143.9 & 188.23486 & 14.69554 & RRab & 0.56958 & 0.78 & 56725.27423 & 20.39 & 0.10 & 70 & 88.4 & LSQ \\
LSQ123426.04+104026.9 & 188.60849 & 10.67415 & RRab & 0.65797 & 0.82 & 55267.27611 & 20.44 & 0.08 & 109 & 94.5 & LSQ \\
LSQ123811.69+110208.9 & 189.54872 & 11.03580 & RRab & 0.63384 & 0.46 & 56778.02524 & 20.35 & 0.06 & 65 & 90.8 & LSQ \\
LSQ123855.41+075409.7 & 189.73087 & 7.90269 & RRab & 0.46390 & 1.45 & 55642.16948 & 20.32 & 0.08 & 119 & 82.8 & LSQ \\
\enddata
\tablenotetext{a}{V-band pulsation amplitude}
\tablenotetext{b}{Heliocentric Julian Date of Maximum Light - 2,400,000.0 }
\tablenotetext{c}{Number of observations used for the light-curve parameter fit}
\tablenotetext{d}{``LSQ" identifies stars detected by LSQ, with parameters derived from LSQ data only.``G+LSQ" identifies additional stars selected from Gaia, with parameters derived from Gaia light curves supplemented by LSQ data when available.}
\tablenotetext{}{Only a portion of this table is shown here to demonstrate its form and content. A machine-readable version of the full table is available.}
\end{deluxetable*}
\endgroup

Figure \ref{fig5} shows there is a small gradient in $\rm R_{gc}$ with position along the OVOD, which is related, of course, to its gradient in $\rm d_{\sun}$.  Subtracting in quadrature the mean $1\,\sigma$ error in $\rm R_{gc}$ (3.1 kpc)from the $1\,\sigma$ scatter in $\rm R_{gc}$, as a function of b, yields 4.3 kpc for the typical thickness of the OVOD in $\rm R_{gc}$, which is small compared to its dimensions in the plane of the sky.  The OVOD appears to be a large and relatively thin cloud of stars, unrelated to the Sgr Stream, which does not have the simple structure of a stellar stream.  The physical properties of its RRLS, which we examine next, provide some insight into its possible origin.
 
\section{The Physical Characteristics of the OVOD RRLS}

The RRLS in dwarf galaxies and globular clusters have distinct properties that depend on their metallicities and to some extent on their horizontal branch morphologies.
This is most clearly demonstrated by the MW globular clusters. On the basis of mean periods of the RRab variables, $\langle \rm P_{ab} \rangle$, these clusters are usually classified as Oosterhoff types I or II (OoI or OoII) if $\langle \rm P_{ab} \rangle$ are $\sim 0\fd55$ or $\sim 0\fd65$, respectively.  Most galactic globular clusters containing 5 or more RRLS fall within one or the other of these two groups, with only a small minority having intermediate values, $0 \fd 58 \leq \langle \rm P_{ab} \rangle \leq 0 \fd 62$, the so-called Oosterhoff intermediate (Oo-int) clusters \citep[e.g.,][]{catelan2009}.  Other characteristics that separate the OoI and OoII clusters are the fraction of cluster RRLS that are RRc variables ($f_c = n_c /(n_{ab}+n_c)$), the mean period of the RRc variables ($\langle \rm P_c \rangle$), the positions of RRab variables in a graph of amplitude vs. logP (the Bailey diagram), metallicity, and horizontal branch (HB) morphology.  In general, the OoII clusters have larger fractions of RRc variables ($\sim 50\%$ vs. $\sim 20\%$ for OoI), slightly longer $\langle \rm P_c \rangle$ ($\sim 0 \fd 37$ vs. $\sim 0 \fd 32$ for OoI), lower [Fe/H], bluer HB's where there is overlap in [Fe/H], and a systematic shift to longer P in the Bailey diagram than do OoI clusters. The origin of this phenomenon appears to be differences in metallicity and the other parameters that produce the differences in HB luminosity and morphology between the OoI and OoII clusters, and consequently, the different directions by which HB stars in the two Oo-types evolve into the instability strip \citep[see][and references therein]{lee1990, catelan2015, li2025}. To some degree, the Oosterhoff classifications can be applied to the satellite galaxies of the MW.  The most luminous and metal-rich of these galaxies, e.g., Fornax and the Magellanic Clouds, contain RRLS that are similar in properties to the ones found in both OoI and OoII clusters, but with a preponderance of OoI ones.  The least luminous galaxies containing RRLS instead have a preponderance of OoII RRLS \citep[e.g.,][]{martinez2017}.

\begin{figure}[h!]
\plotone{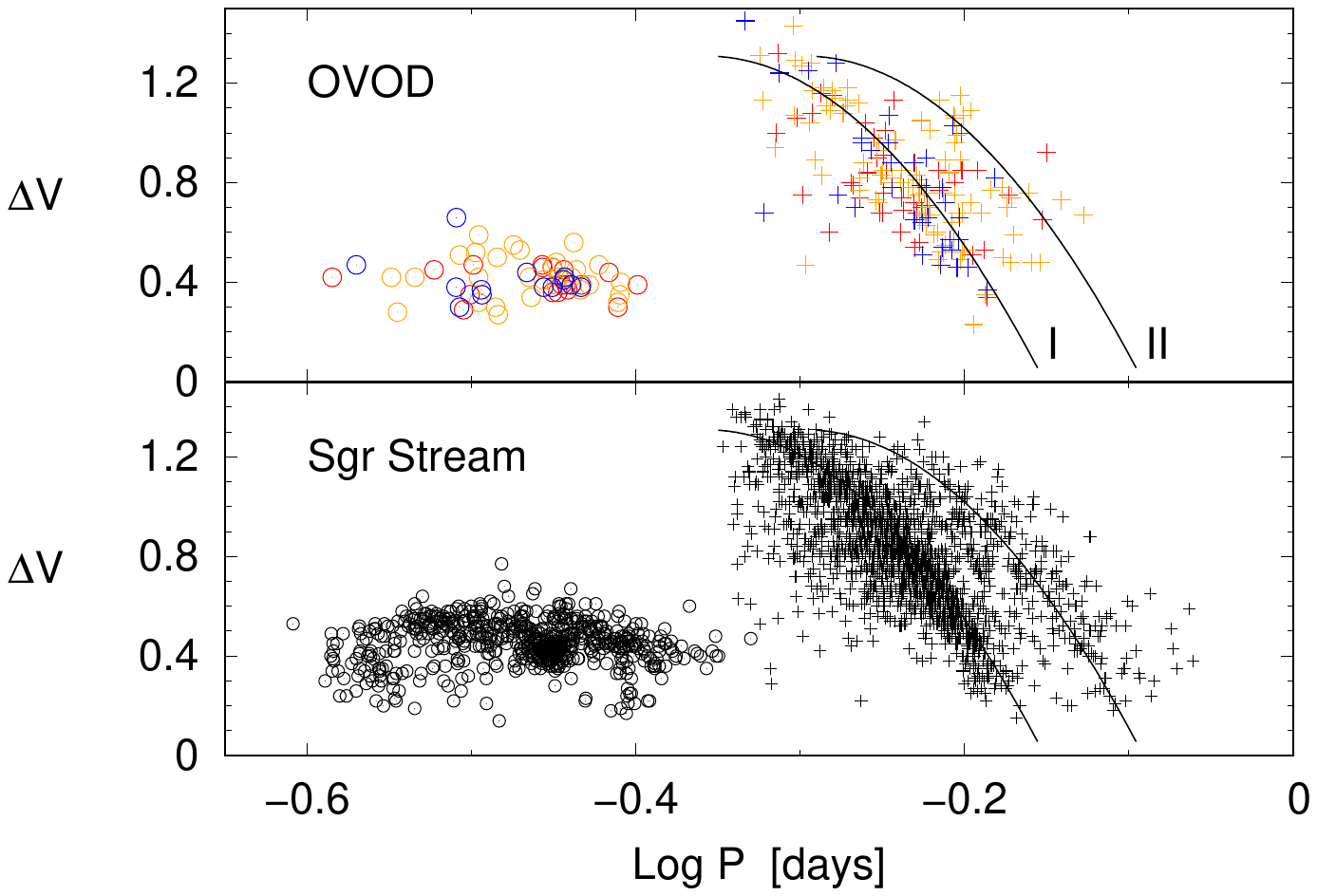}
\epsscale{1.2}
\caption{Using the LSQ sample of RRLS, the V amplitudes of the RRLS pulsations are plotted against the log of their periods (the ``Bailey Diagram").  The RRab and RRc variables are depicted by crosses and open circles, respectively. The top plot is for RRLS in the extended OVOD region (OVOD-C: orange, OVOD-S: red, OVOD-N: blue). The bottom plot is for the Sgr Stream RRLS.  The solid lines show the mean locations of the RRab variables in Oosterhoff I and II globular clusters \citep{zorotovic2010}.}
\label{fig7}
\end{figure}

In Figure \ref{fig7}, we plot the Bailey diagram for the RRLS that were identified in the LSQ survey and lie in the extended OVOD ($3 \,\sigma$ overdensity) region, and for comparison, the RRLS in the Sgr Stream, which we define here as the LSQ RRLS with $40 \leq \rm d_{\sun} \leq 70$ kpc lying within $\pm13^\circ$ of the ridge line of the Stream (i.e., the stars between the dashed lines in the bottom diagram in Fig. \ref{fig3}, restricted to $d_{\sun} \leq 70$ kpc to avoid any overlap with the OVOD).  In Figure \ref{fig7}, the RRLS in the three regions of the OVOD are plotted using different colors and compared with the mean lines of OoI and OoII globular clusters \citep{zorotovic2010}.  One can see from this diagram that the populations of the RRab and RRc variables in the three subdivisions of the OVOD are basically the same. There are also no obvious differences between the OVOD variables and the ones in the Sgr Stream, and the data in Table \ref{table1} indicate this as well.  While this might suggest a connection, this is unlikely in our opinion because of the large differences in $\rm d_{\sun}$ and location on the sky (see above). 

The properties of the RRab variables are arguably the best diagnostic of the Oosterhoff phenomenon.  The values of $\langle \rm P_{ab} \rangle$ for the whole OVOD and the Sgr Stream (see Table \ref{table1}) lie on the border of the Oo-int range.  This intermediate status is clearly not due to large populations of RRab variables lying between the OoI and OoII lines in Figure \ref{fig7}, but rather to a broad mixture of stars.  The majority of the RRab stars fall near the OoI line or to the short period side of it.  A minority lie near the OoII line.  To estimate the percentage of stars that resemble the RRLS in the OoII globular clusters (OoII\% in Table \ref{table1}), we follow \citet{zinn2014} by counting as OoII stars, the RRab that lie to the long period side of a curve that is mid-way between the OoI and OoII lines and compare that to the total number of RRab \footnote{The few RRab with $\Delta \rm V > 1.3$ were excluded.}. There is some variation in OoII \% among the three OVOD samples, with OVOD-C having a higher OoII \% than the other two.  Changing the definition of the OoI and OoII lines to the ones given by \citet{fabrizio2019} changes the OoII\% numbers, but not this trend. This, like the existence of a slight distance gradient, may suggest a more complex origin of the OVOD. Note, however, that other than OoII\%, the properties in Table \ref{table1} for the different OVOD samples are consistent with each other. 

The same Oosterhoff lines and procedure to define OoII\% that were used in construction of Table\ref{table1} were also adopted by \citet{martinez2017} in their analysis of the RRLS populations in the satellite galaxies of the MW and M31.  The OoII\% of the OVOD is similar to that of the Sgr galaxy and many of the other large satellite galaxies of the MW and M31 \citep[see][]{martinez2017}.  The smallest and most metal-poor of satellites have larger OoII\% than the OVOD.  This, together with the large number of RRLS in the OVOD, suggests that it is the debris of a large dwarf galaxy.

\begingroup
\setlength{\tabcolsep}{7pt}
\begin{deluxetable*}{lcccccc}
\tablecaption{Global Properties of the RR Lyrae Variables}
\label{table1}
\tablehead{
\colhead{Object} & \colhead{$n_{ab}$} & \colhead{$n_c$} & \colhead{$\langle \rm P_{ab} \rangle$} & \colhead{$\langle \rm P_c \rangle$} & \colhead{$n_c/(n_{ab}+n_c)$} & \colhead{OoII\%}\\[-0.3cm]
\colhead{} & \colhead{} & \colhead{} & \colhead{(d)} & \colhead{(d)} & \colhead{} & \colhead{}
}
\startdata
OVOD total & 175 & 61 & $0.585\pm0.004$ & $0.343\pm0.004$ & $0.26\pm0.03$ & $19\pm3$\\
OVOD-C & 94 & 30 & $0.590\pm0.006$ & $0.344\pm0.006$ & $0.24\pm0.04$ & $24\pm4$ \\
OVOD-S & 46 & 18 & $0.580\pm0.008$ & $0.347\pm0.008$ & $0.28\pm0.06$ & $16\pm5$ \\
OVOD-N & 35 & 13 & $0.581\pm0.008$ & $0.334\pm0.008$ & $0.27\pm0.06$ & $9\pm5$ \\
Sgr Stream & 1983 & 645 & $0.584\pm0.002$ & $0.342\pm0.002$ & $0.25\pm0.01$ & $18\pm1$
\enddata
\end{deluxetable*}
\endgroup

Support for this hypothesis is also provided by the metallicities of the RRab, which can be estimated using the logP-[Fe/H] relation from \citet{sarajedini2006} \citep[on the scale of][]{zinn1984}.  While not of high precision \citep[see also][]{sarajedini2023}, this relation should yield a good estimate of the mean. For the 175 RRab in the OVOD, we obtain a mean [Fe/H] of $-1.60\pm0.02$.  The much less certain [Fe/H] estimates for the individual stars range from -2.4 to -0.8, with a standard deviation of $\sigma = 0.32$ about the mean [Fe/H].  The mean and range are indicative of a large dSph galaxy.  For example, with the same logP-[Fe/H] relation, the LSQ observations of 209 previously known RRab in the Sculptor dSph galaxy yield a mean [Fe/H] = $-1.69\pm0.03$ and a range from -3.0 to -1.0 ($\sigma = 0.43$), which are close to the values obtained from spectroscopy of red giants \citep[see][and refs. therein]{tolstoy2023}. After Fornax and LeoI, Sculptor is the $3^{rd}$ most luminous intact dSph satellite of the MW \citep[$\rm M_V = -11.1$,][]{mcconnachie2012}.

\citet{chandra2023} have provided evidence that the OVOD is debris from the GSE merger (see below).  The mean [Fe/H] obtained above for the OVOD is essentially identical to the values obtained from samples of less distant RRLS that are likely GSE members based on their kinematics.  \citet{zinn2020} obtained $-1.58\pm0.03$ from spectroscopic measurements of 91 RRLS, and \citet{muraveva2026} obtained $-1.57\pm0.25$ from the Fourier parameters of the light curves of 315 RRLS observed by Gaia. If the OVOD is not GSE debris, then the properties of its RRLS suggest an origin in another relatively large galaxy.


\section{Is the OVOD a Stream or a Shell?}

The difference between stream and shell geometries is quite important for understanding the properties of a merger event.  Simulations \citep[e.g.,][]{hendel2015} show that a minor-merger along an orbit of low eccentricity causes the stripped stars to spread out approximately along the orbit of the progenitor, producing a stream. If the orbit has high eccentricity, the stars spread out along the orbit, but also perpendicular to it at apocenter, producing a shell.  

The morphology of the OVOD is reminiscent of some of the distant halo substructures found by the PAandAS survey of M31's halo \citep{mcconnachie2018}.  Perhaps the best examples for comparison are the so-called East and Southwest Clouds.  Similar to the OVOD, these overdensities are large in size, lie at large distances from the center of M31 ($\gtrsim 80$ kpc in projection), and their longest dimension is roughly perpendicular to the radial direction to M31.  \citet{mcconnachie2018} have suggested that these M31 substructures, and in addition Stream A, may be shells produced by the merger of a relatively large galaxy.  While the OVOD may have been formed by the same type of merger event that formed these M31 clouds, it was probably on a smaller scale.  The estimated luminosities of the East and Southwest clouds are large \citep[$\rm M_V = -10.7$ and $-12.1$ respectively,][and refs. therein]{mcconnachie2018}, and bracket that of the Sculptor dSph satellite of the MW \citep[$-11.1\pm0.5$,][]{mcconnachie2012}, which contains $\sim 3$ times the number of RRLS in the OVOD. Moreover, these M31 clouds contain a few globular clusters each \citep{mackey2019, mcgill2025}, whereas none of the known MW globular clusters lie in or near the OVOD.

Since the discovery of the OVOD by \citet{sesar2017}, there have been several suggestions in the literature about its possible connection to the VOD and other halo overdensities \citep[e.g., the Cocytos Stream;][]{aganze2025}.  We focus our attention here on the analysis by \citet{chandra2023} because it rests on observations of red giants in the OVOD rather than extrapolations from observations made at smaller distances and at other lines of sight in the halo.  They suggested that the OVOD and part of the Pisces overdensity are shells produced by the GSE merger.

\citet{chandra2023} selected for spectroscopy an unbiased sample of 13 red giants (``Mike-N'' sample) in the OVOD, which have $\rm d_{\sun}$ in the range of $\sim 70-100$ kpc.  They also observed a southern galactic latitude sample of 17 red giants near the Pisces overdensity, which were selected on the basis that their proper motions suggested retrograde orbits.  Their analysis showed that the majorities of both samples of stars have the space velocities indicative of similar retrograde orbits, and consequently they argued that the OVOD sample and their southern sample were apocenter pileups of the orbit of a disrupted satellite galaxy.  Their suggested orbit was in rough agreement with the orbit that \citet{naidu2021} predicted from simulations of the GSE merger event, and consequently, \citet{chandra2023} argued that they had found a stream of stars that were stripped from the GSE progenitor before it lost angular momentum and its orbit became radial.  The relatively high metallicities that they measured from their spectra are also consistent with their hypothesis.

\begin{figure}[h!]
\epsscale{1.1}
\plotone{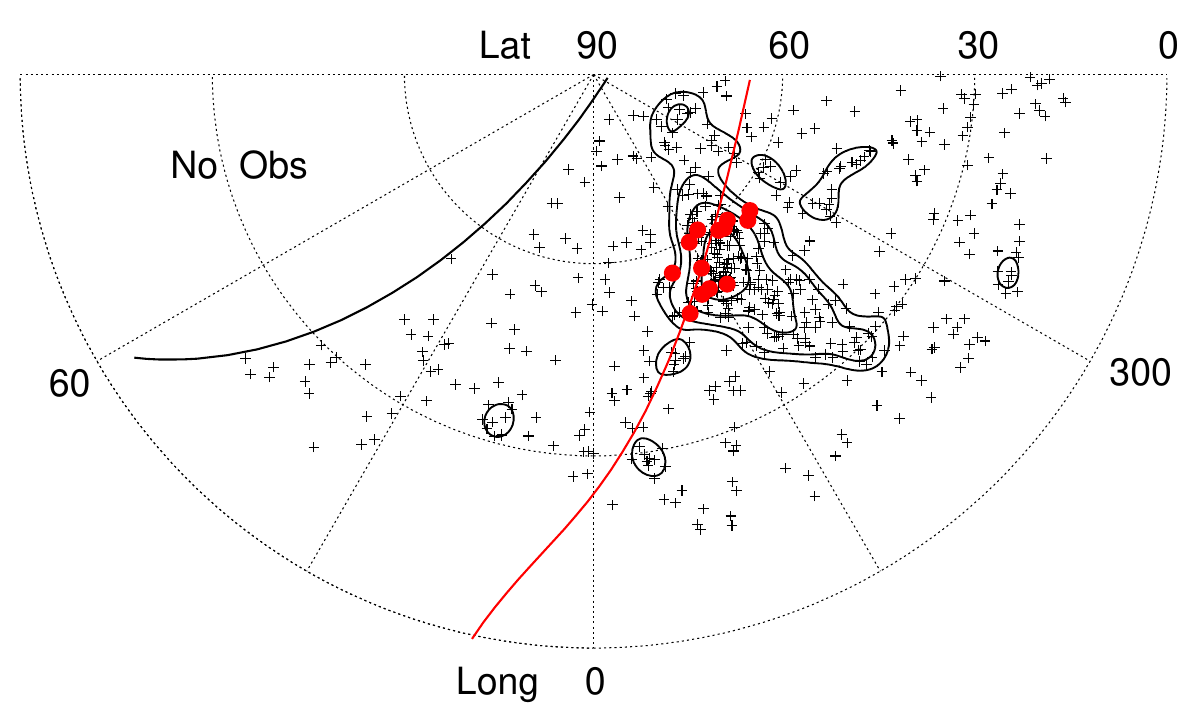}
\caption{The \textit{red curve} is the great-circle approximation that \citet{chandra2023} made to the orbits of their ``Mike-N'' sample of red giants (\textit{red points}).  The \textit{black crosses} are the LSQ and Gaia RRLS in the distance range $75 \leq \rm d_{\sun} \leq 100$ kpc (black crosses).  The contours are the same contour levels plotted in Figure 6.}
\label{fig6}
\end{figure}

To compare with our results, we have measured from the red curve in Fig. 6 of \citet{chandra2023} the galactic coordinates of the great circle that they fit to the orbits that they calculated for their sample of red giants. While our coordinates are rough, they capture to within a few degrees the path of the GSE near the OVOD (the red curve in our Figure \ref{fig6}), as inferred by \citet{chandra2023}. As expected, this curve passes through the sample of ``Mike-N'' red giants that \citet{chandra2023} used to infer the orbit. It roughly aligns with the bulge in the density contours of the OVOD RRLS. There are several RRLS outside the OVOD that align with the red curve in Figure \ref{fig6}, including two small overdensities enclosed by $3\, \sigma$ contours.  

If the whole OVOD were part of the stellar stream identified by \citet{chandra2023}, then one would expect its long dimension to be aligned with the red curve in Figure \ref{fig6}, which is not the case. Figure \ref{fig6} shows that as the red curve passes through the OVOD, it is at angles of $\sim 25^\circ$ and $\sim 70^\circ$ with respect to the northern and southern ``legs'' of the OVOD, respectively. The large angle with the southern ``leg'' appears to rule out a stream morphology.  The alternative is that it is a shell that was produced by a merger on a highly eccentric orbit, which is consistent with the hypothesis that the OVOD is GSE debris \citep[][]{chandra2023}.  The measurement of the orbits for more stars in the OVOD, particularly for stars in OVOD-S and OVOD-N, and theoretical simulations are needed before one can be sure of this interpretation. 

\section{Summary}

The LSQ survey has shown that the previous detection of the OVOD by \citet{sesar2017} was only of its densest region, and that it extends $\sim 20^\circ$ north and south of its previously known dimensions. It is $\sim 57^\circ$ ($\sim 85$ kpc) long and from $\sim 10^\circ$ ($\sim 15$ kpc) to $\sim 35^\circ$ ($\sim 50$ kpc) wide.  There is a bend at its densest part of $\sim 35^\circ$ with respect to a great circle.  Its mean $\rm d_{\sun}$ and $\rm R_{gc}$ are 85 and 82 kpc, respectively, and in both quantities there is a small gradient with position in the sense that the southern part is closer than the northern.  Although spatially extended on the sky, the overall OVOD is relatively thin, $\sigma_{\rm R_{gc}} = 4.3$ kpc, after correction for observational error.  Overall, the OVOD resembles a cloud similar to the ones found in the outer halo of M31 in that its long dimension is approximately perpendicular to the direction to the Galactic center.

As is the case with most of the dSph satellite galaxies of the MW and M31, the OVOD RRLS have a range of properties that do not fit either the OoI or OoII classifications that fit the majority of MW globular clusters.  The properties of the OVOD RRLS closely match those of the Sgr Stream, but we rule out a connection because the OVOD deviates by as much as $30^\circ$ from the Stream's plane and is $\gtrsim 20$ kpc more distant.  The low fraction of OVOD RRab that lie near the locus of the OoII RRab in the Bailey diagram, and the mean [Fe/H] of -1.60 of the RRab, both point to an origin in another large dwarf progenitor, which was similar to the largest satellite galaxies of the MW.

Our comparison with the results of \citet{chandra2023}, who proposed that the OVOD is a shell of stars from the GSE merger, revealed that their proposed orbit coincides with the bulge in the OVOD. However, the whole OVOD does not align with their orbit.  The southern part of the OVOD makes an angle of $\sim 70^\circ$ to their orbit, and it extends roughly $30^\circ$ from it. This supports the hypothesis that it is a shell rather than a stellar stream.    

The overall picture of the OVOD fits the hypothesis that it is debris from a large dwarf galaxy that collided with the MW on a near radial orbit.  However, there is clearly a need for more space velocities for OVOD stars, particularly for its southern extension, which provides the strongest evidence against stream morphology.  If indeed the OVOD is a shell, then its large $\rm R_{gc}$ may be a powerful constraint on possible models.  Since other halo substructures such as the VOD and the HAC are suspected to be shells of a radial-orbit merger \citep[either the VRM or the GSE, e.g.,][]{donlon2020, balbinot2021}, it may be that one merger produced interleaved shells or that there have been multiple mergers on high-eccentricity orbits \citep[e.g.,][]{donlon2022}.  This is ripe for modeling.  Since \citet{vasiliev2021} found
that including the LMC in their models improved the fit to the observations of the leading Sgr Stream, which lies in the foreground of the OVOD in the same region of the sky, this may also be necessary for models of the OVOD.

Finally, we have assumed above that the extended OVOD we have found is a single halo overdensity, with one origin. It is plausible that the bend in the density contours and the small gradient in $\rm R_{gc}$ are signs that this picture is oversimplified. Obviously, further investigation of the OVOD is warranted, and it is almost certain to reveal more evidence of the chaotic past of the MW.    

\section{Acknowledgments}

Reprocessing of the La Silla QUEST data to produce the improved RR Lyrae catalog used here was enabled by essential support from National Science Foundation grant AST-1909101.  The construction and execution of the La Silla QUEST survey itself was supported by the Department of Energy and Yale University.

This work has also made use of data from the European Space Agency (ESA) mission
{\it Gaia} (\url{https://www.cosmos.esa.int/gaia}), processed by the {\it Gaia}
Data Processing and Analysis Consortium (DPAC,
\url{https://www.cosmos.esa.int/web/gaia/dpac/consortium}). Funding for the DPAC
has been provided by national institutions, in particular the institutions
participating in the {\it Gaia} Multilateral Agreement.

The text of this paper was proofread with assistance from ChatGPT.

\appendix
\section{The $\rm M_{V}-Log\, P$ Relations}
It is well documented by both observation and theory that the luminosities of RRLS are functions of their periods, colors, and metallicities.  Because our survey consists of observations solely in the V pass-band, relationships involving color cannot be applied.  In principle, we could apply an $\rm M_{V} - [Fe/H]$ relationship, after determining the [Fe/H] values of the variables from Fourier decomposition of their light curves (e.g., the $\phi_{31} - \rm [Fe/H]$ relation).  However, for $d_{\sun}$ $ \gtrsim 70$ kpc, the crucial range for studying the OVOD, many of our light curves for RRLS do not produce reliable measures of the Fourier parameters. In \citet{zinn2014}, we assumed one value of $\rm M_{V}$ for all RRLS, which for the metallicity range of the RRLS in the galactic halo yields relative errors ($\sigma_{d}/d,\,  \sigma = \rm std.\,dev.$) of $\sim 0.07$ \citep{vivas2006}. Since metal-poor RRLS are on average more luminous than metal-rich ones, the adoption of one value for $\rm M_V$ introduces a bias by [Fe/H] in the distance measurements. We show below that $\rm M_{V} - logP$ relationships, separately for RRab and RRc variables, yield smaller relative errors, and at most a small [Fe/H] bias.  

In a stellar population of one age and one [Fe/H], which is at least approximately true in the case of most galactic globular clusters, the RRLS have a small range in mass.  Consequently, their periods depend almost entirely on their mean effective temperatures ($\rm T_{e}$) and luminosities.  The division of the instability strip by $\rm T_{e}$ into zones occupied separately by RRab and RRc variables restricts, but does not eliminate, the period variation at constant luminosity among the RRab and RRc variables.  Evolution away from the zero-age horizontal branch (HB) produces a range in luminosity among the RRLS, which is manifested in the thickness of the HB in luminosity. In individual globular clusters, $\langle \rm V \rangle$ correlates with the periods of the  RRab and RRc variables with scatter due to the range of mean $\rm T_e$.  It is less clear that the RRLS in the galactic halo should also exhibit a tight $\rm M_{V} - logP$ relation, because they span a wide range in [Fe/H] and HB luminosity is a function of [Fe/H].  Undoubtedly, there is some range in age as well. But $\rm M_{V}$ is a function of [Fe/H], and at least among the RRab variables, logP can predict [Fe/H], albeit not with high precision \citep[][]{sarajedini2023}. By constructing  $\rm M_V - logP$ relations of several globular clusters of different [Fe/H], we show that using logP alone is still useful for improving the determination of $\rm M_V$.           

\begin{figure}[h!]
\plotone{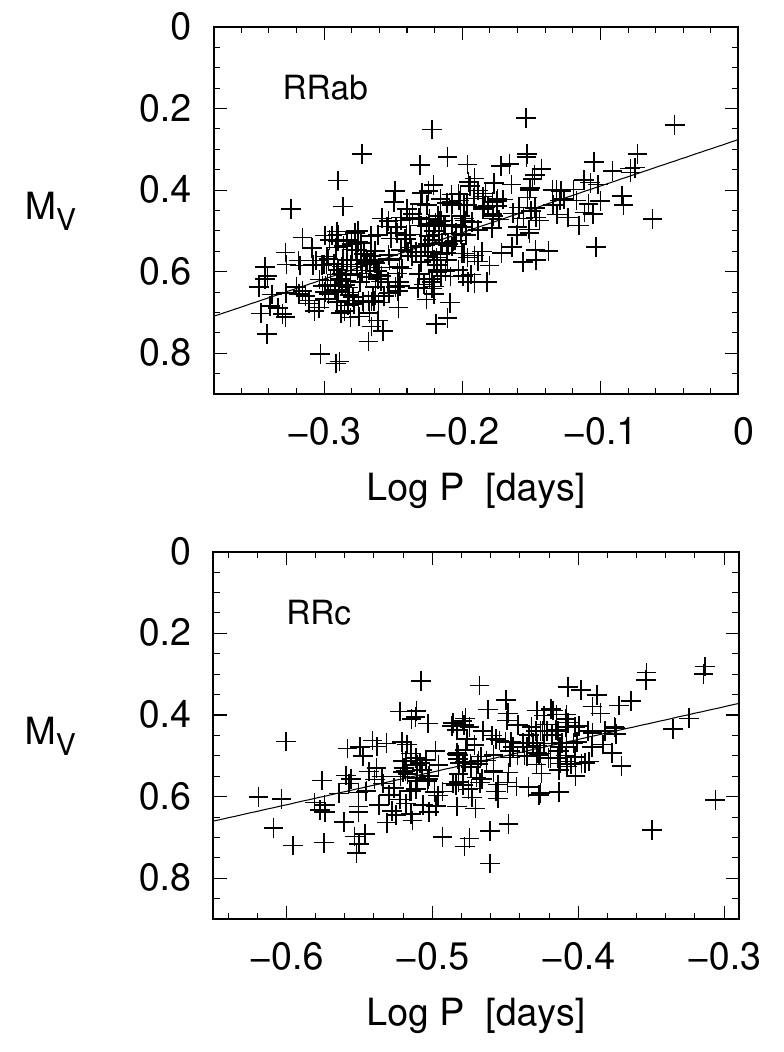}
\caption{$\rm M_{V}$ is plotted against log P for the 347 RRab and 233 RRc variables in 9 globular clusters with well-determined distances from \citet{baumgardt2021}. The solid lines are least-squares fits to these data (Equations (A1) and (A2)), which are used in the distance measurements.}
\label{figA1}
\end{figure}

Recently, \citet[][hereafter B\&V]{baumgardt2021} derived precise distances to globular clusters by averaging values from the literature and their own analysis.  From their compilation, we selected 9 globular clusters (NGC 4590 (M68), 5024 (M53), 5139 ($\omega \, \rm Cen$), 5272 (M3), 5466, 5904 (M5), 6341 (M92), 6362, 7089 (M2)) that have the following attributes: relative distance errors $\lesssim 0.01$, at least 10 RRLS, and small and uniform interstellar reddenings (E(B-V) $\le 0.12$, Harris 1996, 2010 ed.). The [Fe/H] of these clusters range from -1.07 to -2.31 \citep{carretta2009}, almost covering the entire range observed in samples of halo RRLS \citep[e.g.][]{liu2022}. NGC 7078 (M15) also meets these criteria, but we found that its RRLS were offset from the other clusters of similar [Fe/H] in Figure \ref{figA1}, and it was therefore excluded.  We purposely included $\omega$ Cen in this sample because, unlike the other clusters, its RRLS span a wide range in [Fe/H]  \citep[-2.58 to -0.85,][]{magurno2019}.  Its RRLS are not offset from the ones in the other clusters.

From the ``Catalogue of Variable Stars in Galactic Globular Clusters'' (CVSGGC)\footnote{https://www.astro.utoronto.ca/gccat/index.html} \citep[see also][]{clement2001}, we downloaded the periods and $\langle\rm V\rangle$ of the type RRab and RRc variables in the above listed clusters.  We only selected stars that had m1 membership flags from \citet{prudil2024}, indicating a probability of membership $\ge0.85$ on the basis of the Gaia proper motions. In every cluster used, the majority of the listed RRLS met this membership criterion. Stars with uncertain periods and $\langle\rm V\rangle$ were excluded, as were stars listed as blends.  Using the distances given by B\&V and the values of E(B-V) by \citet[][2010 ed.]{harris1996}, and assuming the interstellar extinction, $\rm A_V = 3.1E(B-V)$, we calculated the absolute V magnitudes ($\rm M_{V}$) of the variables, which are plotted against logarithms of their periods in Figure \ref{figA1}.  The lines are least-squares fits to the data, and their equations are listed below.
\begin{equation}
\rm M_{V}^{ab} = 0.276(\pm 0.017)-1.140(\pm 0.070)\rm logP
\end{equation}

\begin{equation}
\rm M_{V}^{c} =0.140(\pm0.039)-0.800(\pm0.083)\rm logP
\end{equation}

The residuals in $\rm M_{V}$ about the lines in Figure  \ref{figA1} have standard deviations of 0.080 and 0.077 for the RRab and RRc stars, respectively. The scatter is due to the variation in mean $\rm T_e$ mentioned above and the errors in $\langle \rm V \rangle$ of the RRLS, the distances, and the $\rm A_V$ values of the clusters.      

The intrinsic dispersions of Equations A1 and A2 compare favorably with some of the more common ways of inferring the luminosities of RRLS and hence their distances.  We consider, for example, several of the different relations that \citet[][]{garofalo2022} obtained using Gaia Early Data Release 3 parallaxes. For 366 RRLS in 15 globular clusters they obtained a $\rm M_V - [Fe/H]$ relation with $\sigma = 0.11$.  They quote $\sigma = 0.02$ for the $\rm M_V - [Fe/H]$ relation that they obtained for 274 field RRLS, which must be barely larger than the errors in the $\langle \rm V \rangle$ measurements of the variables.  For 190 stars in their field RRLS sample, they also derived the period-metallicity-Wesenheit G magnitude relation, which has the advantage of being reddening independent. This relation, which includes both log P and [Fe/H], has $\sigma = 0.09$.

To test the usefulness of Equations A1 and A2 for measuring distances of RRLS in the LSQ survey, we cross-matched the survey with the CVSGGC, and in addition, with the catalogs of RRLS in 4 dSph galaxies: Bootes I \citep{siegel2006}, Crater II \citep{vivas2020}, Sculptor \citep{martinez2016}, and Sextans \citep{vivas2019}. The combination of several dead CCDs in the LSQ camera and the gaps between the live CCDs conspired to limit the coverage of these objects.  Because our survey employs aperture photometry, the photometry of some of the detected RRLS in crowded fields had to be discarded because of blending. Reliable measurements of $\langle \rm V \rangle$ were obtained for 107 RRLS in 18 globular clusters and for 646 in the 4 galaxies. Using these data, Equations A1 and A2, and the interstellar extinctions from the dust maps, we calculated the distance moduli, $\rm (m-M)_0$, of the RRLS and the mean values for each of these objects.  The results are presented in Table \ref{tableA1}, where for the globular clusters the values and errors of $\rm (m-M)_0$ were calculated from the distances and errors given by B\&V. Because the distance that B\&V list for Pal 3 is based on only 2 literature values from color-magnitude diagram (CMD) fitting \footnote{ https://people.smp.uq.edu.au/HolgerBaumgardt/globular/fits/pal3.html}, its value of $\rm (m-M)_0$ is particularly uncertain.  The LSQ value of $\rm (m-M)_0$ agrees well with the shorter of the two distance estimates from CMD fitting, but it is more than $1\, \sigma$ from the mean.  For the galaxies, we adopted the values listed by \citet{battaglia2022}, with the exception of Sextans for which they list $19.64\pm0.01$.  This unrealistically small error appears to be the precision that \citet{vivas2019} obtained using a period-luminosity-metallicity relation with their g-band light curves of Sextans RRab variables. With their r-band light curves and a similar relation, they obtained $19.677\pm0.008$.  We have adopted here $19.66\pm0.05$, which may be an underestimate of the true uncertainty \citep[for example,][list $19.67\pm0.1$]{mcconnachie2020}.  The $\rm A_V$ in Table \ref{tableA1} is the average of individual ones for the RRLS in the objects, which in all cases varied by very small amounts. The error in the LSQ $\rm (m-M)_0$ in Table \ref{tableA1} is the std. error of the mean of the individual values if the number of RRLS, $\rm N_{RR}$, $\geq$ 2. For some clusters, this error is smaller than $\sim 0.08/\rm N_{RR}^{\frac{1}{2}}$, and  therefore probably an underestimate. For the clusters with $\rm N_{RR} = 1$, the error is the $\pm 1\, \sigma$ confidence interval on $\rm M_V$ read from Equation A1 or A2. Table \ref{tableA1} lists the [Fe/H] of the globular clusters \citep{carretta2009} and the \textit{mean} [Fe/H] of the galaxies \citep[for Sculptor,][]{tolstoy2023}\citep[all others][]{battaglia2022}.

\begin{deluxetable}{lclcrcr}
\tablecaption{Comparison of Distance Moduli}
\label{tableA1}
\tablehead{
NGC/Name & [Fe/H] & $\rm (m-M)_0$ lit. & $\rm A_V$ & $\rm N_{RR}$ & $\rm (m-M)_0$ LSQ & $\Delta \rm (m-M)_0^\tablenotemark{a}$}
\startdata
1904  &   -1.58 &   15.583$\pm 0.030$ &   0.083 &  1 &  15.554$\pm 0.077$ &  -0.028$\pm 0.083$ \\
Pal 3  &   -1.67 &   19.885$\pm 0.074$ &   0.120 &  4 &  19.759$\pm 0.006$ &  -0.126$\pm 0.074$ \\
4147  &   -1.78 &   16.340$\pm 0.025$ &   0.071 &  4 &  16.380$\pm 0.048$ &   0.040$\pm 0.054$ \\
4590  &   -2.27 &   15.086$\pm 0.021$ &   0.167 & 11 &  15.027$\pm 0.016$ &  -0.059$\pm 0.026$ \\
5024  &   -2.06 &   16.336$\pm 0.022$ &   0.057 & 21 &  16.306$\pm 0.012$ &  -0.030$\pm 0.025$ \\
5053  &   -2.30 &   16.220$\pm 0.029$ &   0.046 &  6 &  16.137$\pm 0.054$ &  -0.083$\pm 0.061$ \\
5634  &   -1.93 &   17.071$\pm 0.052$ &   0.155 &  4 &  17.047$\pm 0.092$ &  -0.024$\pm 0.106$ \\
Pal 5  &   -1.41 &   16.706$\pm 0.051$ &   0.153 &  4 &  16.752$\pm 0.018$ &   0.046$\pm 0.054$ \\
5897  &   -1.90 &   15.493$\pm 0.041$ &   0.379 &  1 &  15.444$\pm 0.081$ &  -0.049$\pm 0.090$ \\
5904  &   -1.33 &   14.369$\pm 0.017$ &   0.105 & 24 &  14.422$\pm 0.015$ &   0.053$\pm 0.023$ \\
6426  &   -2.37 &   16.581$\pm 0.037$ &   0.925 &  5 &  16.691$\pm 0.025$ &   0.110$\pm 0.044$ \\
6809  &   -1.93 &   13.641$\pm 0.021$ &   0.373 &  1 &  13.478$\pm 0.077$ &  -0.163$\pm 0.080$ \\
6864  &   -1.29 &   16.561$\pm 0.048$ &   0.431 &  1 &  16.717$\pm 0.078$ &   0.157$\pm 0.092$ \\
6981  &   -1.48 &   16.109$\pm 0.024$ &   0.159 & 10 &  16.195$\pm 0.016$ &   0.086$\pm 0.029$ \\
7089  &   -1.66 &   15.340$\pm 0.021$ &   0.131 &  6 &  15.347$\pm 0.021$ &   0.007$\pm 0.030$ \\
7099  &   -2.33 &   14.636$\pm 0.023$ &   0.143 &  2 &  14.531$\pm 0.040$ &  -0.106$\pm 0.046$ \\
Pal 13 &   -1.78 &   16.853$\pm 0.037$ &   0.303 &  1 &  16.836$\pm 0.080$ &  -0.017$\pm 0.088$ \\
7492  &   -1.69 &   16.936$\pm 0.051$ &   0.099 &  1 &  16.839$\pm 0.081$ &  -0.097$\pm 0.096$ \\
Bo\"otes I  & -2.55 &   19.11$\pm 0.08$  &   0.048 & 11 &  19.044$\pm 0.016$ &  -0.066$\pm 0.082$ \\
Crater II & -1.98 &   20.33$\pm 0.07$  &   0.094 & 78 &  20.314$\pm 0.009$ &  -0.016$\pm 0.071$ \\
Sculptor & -1.82 &   19.62$\pm 0.04$  &   0.051 & 373 & 19.616$\pm 0.005$ &  -0.004$\pm 0.040$ \\ 
Sextans  & -1.90 &   19.66$\pm 0.05$  &   0.126  & 184 & 19.713$\pm0.006$ &   0.053$\pm0.050$
\enddata
\tablenotetext{a}{LSQ-lit, errors combined in quadrature.}
\end{deluxetable}

\begin{figure}[h!]
\plotone{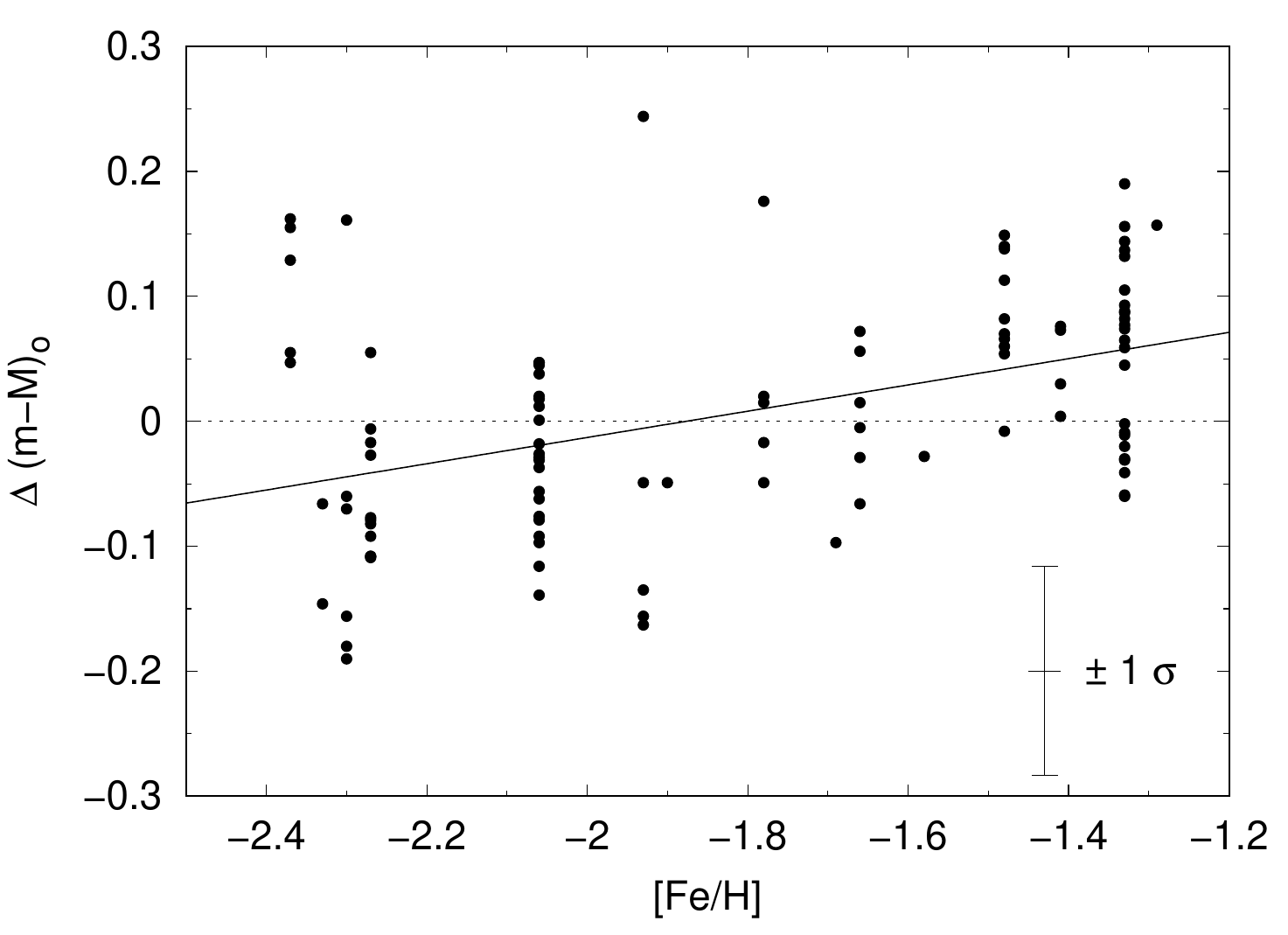}
\caption{For the individual RRLS in 17 globular clusters, $\Delta \rm (m-M)_{0}$, the difference between the distance modulus based on the $\rm M_{V} - logP$ relations and that derived from the distance measured by \citet{baumgardt2021}, is plotted against the [Fe/H] of the cluster \citep{carretta2009}. The solid line is a least-squares fit to the data.  The error bar is $\pm 0.08$, the residuals from Equations (A1) and (A2).}
\label{figA2}
\end{figure}

For the RRLS in 17 of the globular clusters (Pal 3 excluded, see above), Figure \ref{figA2} plots the difference $\Delta \rm (m-M)_0$, in the sense LSQ - B\&V, against [Fe/H], assuming each star has the [Fe/H] of its cluster. The mean $\Delta \rm (m-M)_0$ is 0.008 with $\sigma = 0.09$, which is consistent with the errors of the individual points. There is, however, a trend with [Fe/H], which is illustrated by a linear least-squares fit to the data (see Figure \ref{figA2}).  It is statistically significant, but its magnitude is very small.  At $\rm [Fe/H] = -2.5$ and $ -1.2$, $\Delta \rm (m-M)_0 \sim -0.07$ and $+0.07$, respectively. Because $\sigma_d /\rm d_{\sun} = 0.46 \sigma \rm (m-M)_0$, the likely systematic error in $\rm d_{\sun}$ is only $\sim 3\%$, and only for the relatively few RRLS occupying the metal-poor and metal-rich tails of the halo [Fe/H] distribution.

\begin{figure}[h!]
\plotone{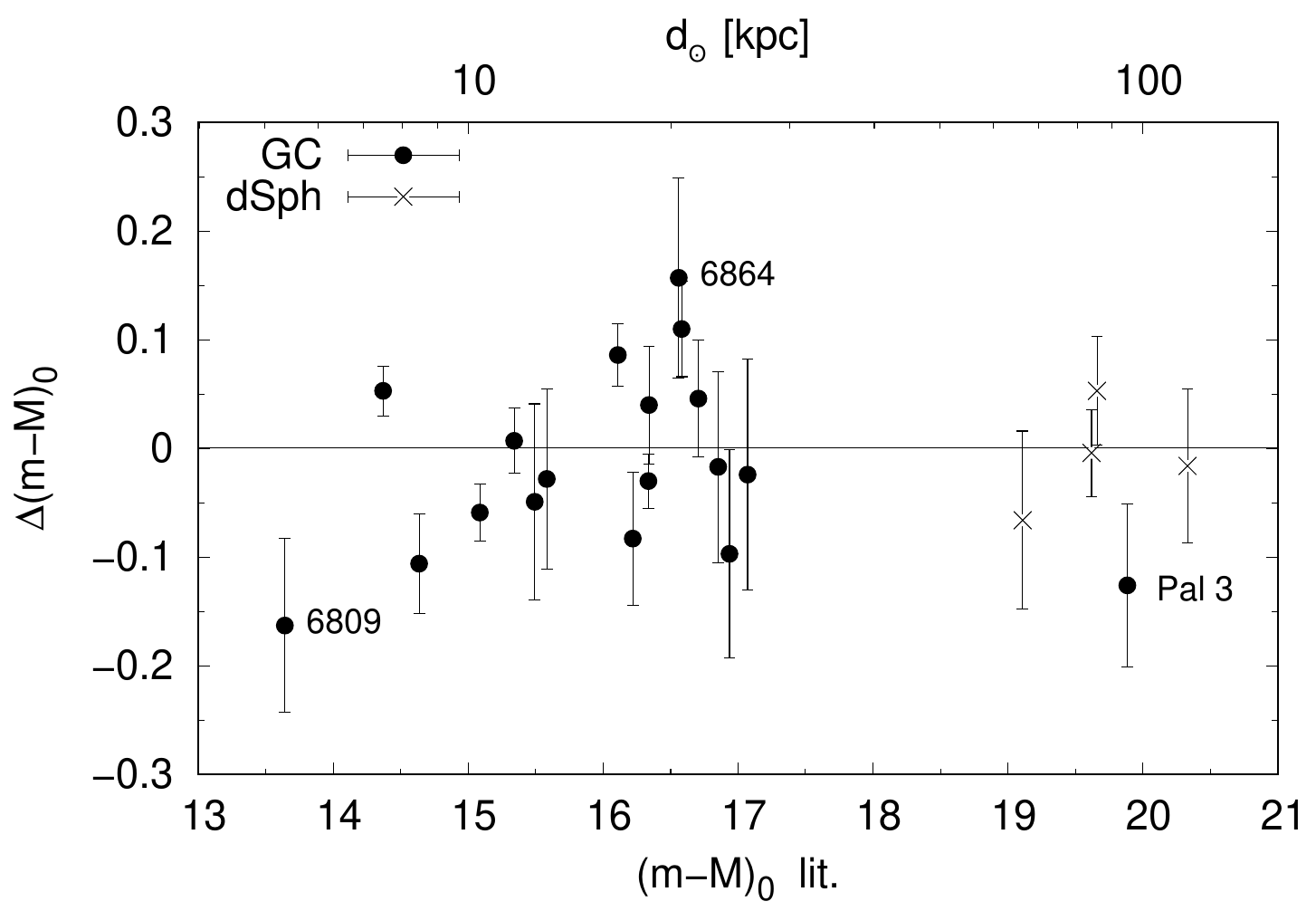}
\caption{For 18 globular clusters and 4 dSph galaxies, $\Delta \rm (m-M)_{0}$ is plotted against $\rm (m-M)_{0}$ (see Table \ref{tableA1}).  The top scale is distance from the Sun ($d_{\sun}$).}
\label{figA3}
\end{figure}

In Figure \ref{figA3}, $\Delta \rm (m-M)_0$ is plotted against $\rm (m-M)_0$ for the globular clusters and galaxies in Table \ref{tableA1}. The 3 objects deviating most from $\Delta \rm (m-M)_0 = 0$ are labeled, which besides Pal 3 (see above) are NGC 6809 and 6864, whose LSQ $\rm (m-M)_0$ values are based on 1 RRLS in the cluster.  Only 1 RRLS was measured in 4 other clusters, and the mean $\Delta \rm (m-M)_0$ for the 6 clusters is $-0.033$ with $\sigma = 0.11$, which corresponds to a relative error in distance of $5\%$. The mean $\Delta \rm (m-M)_0$, weighted by $1/\sigma^2$, of all 22 objects in Table \ref{tableA1} is 0.003 with $\sigma = 0.066$.  Because the spectroscopy of the red giants in the 4 galaxies reveals that each has a wide internal range in [Fe/H], their RRLS populations are probably also heterogeneous in [Fe/H].  The LSQ values of $\rm (m-M)_0$ for the samples of RRLS in the Bo\"otes I, Crater II, Sculptor, and Sextans galaxies have $\sigma$'s of 0.053, 0.082, 0.090, and 0.075, respectively.  These $\sigma$'s include contributions from the errors in the LSQ measurements of $\langle \rm V \rangle$, the uncertainties in Equations A1 and A2, and the sensitivity of the method to [Fe/H] differences.  They suggest that over the range $70 \lesssim \rm d_{\sun} \lesssim 120 $ kpc, the method outlined here yields relative errors of $\sim 4\%$. To be conservative, we have adopted 5\% in the analysis of the LSQ observations of the OVOD (see main text).      

\section{RR Lyrae Stars in Gaia DR3}

Gaia DR3 \citep{GaiaDR3_overview} represents a significant improvement in the detection and characterization of RRLS compared to DR2, with the DR3 vari\_rrylrae table \citep{clementini2023} containing many fewer spurious objects and classifications. Since the nominal Gaia sensitivity covers the range of apparent brightness expected for RRLS in the distance range of interest for this paper, 75-100 kpc, and the Gaia DR3 survey has a potentially different selection function from LSQ, it is useful to compare and then, if possible, merge the RRL samples from both surveys.  The two surveys provide an important cross-check on each other's purity and completeness, and objects that are ambiguous in one survey may not be in the other. LSQ, for example, tends to have significantly more  but lower accuracy observations of a given object (often $>100$) while Gaia DR3 tends to provide fewer ($\sim 30$)  but more accurate observations. Both surveys can have significant, but fortunately different, systematic photometry problems. Also, because the Gaia scan pattern does not suffer from the day-night exposure problem of a ground-based telescope, the Gaia data can break period degeneracies caused by the day-related period aliases that plague ground-based data.

The purity and completeness of the LSQ RRLS sample have already been discussed in \citet{coppi2024} on the basis of comparisons with other deeper surveys of RRLS in dwarf galaxies. For regions with good spatial and temporal coverage, such as most of the OVOD discussed here, we think the LSQ completeness should be better than 90\% out to 100 kpc, and the purity should be better than 95\% due to the conservative selection criteria we applied. A first comparison to the Gaia DR3 sample, however, reveals a significant discrepancy in the number and type of RRLS found at fainter magnitudes. In particular, if we define the Gaia DR3 RRc as those stars having a period entry only in the first overtone (p1o) column of the DR3 RRLS table, then at average magnitude $\langle G\rangle >20$ and galactic latitude $|b|>15^\circ$ and declination $\delta<25^\circ$ (the sky covered by LSQ), we find 2617 RRc. The corresponding number of $\langle G\rangle >20$ RRab (defined as those having only a period entry in the fundamental period column) is 3092, for an RRab-to-RRc number ratio of 1.18. This stands in sharp contrast to the $\sim$ 3:1 number ratio expected for a halo population, and what LSQ found.
The spatial distribution of the faint DR3 RRc stars
is also unusual, with an increasing concentration towards the Galactic plane and bulge not seen by LSQ. This distribution, however, does resemble that found by an automated LSQ search that relied mostly on finding a strong periodicity signal, with little weight given to the actual light curve shape, period, or intrinsic average color. When we examined the light curves of those (automated search) objects in detail, they usually turned out to be real variables that were not RRLS (often red contact binaries whose density increases sharply as one goes towards the Galactic plane) or badly photometered galaxies. To see if this was happening in the Gaia DR3 sample, we inserted the Gaia light curves into our final (manual) candidate inspection tool for LSQ, treating Gaia candidates exactly as we would LSQ candidates. Many of the Gaia candidates that did not have LSQ RRLS counterparts did indeed turn out to be clearly bad photometry (often but not exclusively associated with extended objects) or variables that were too red and did not have the right light curve shapes or periods to be RRLS. 

\begin{figure}
    \centering
    \includegraphics[width=0.43\linewidth]{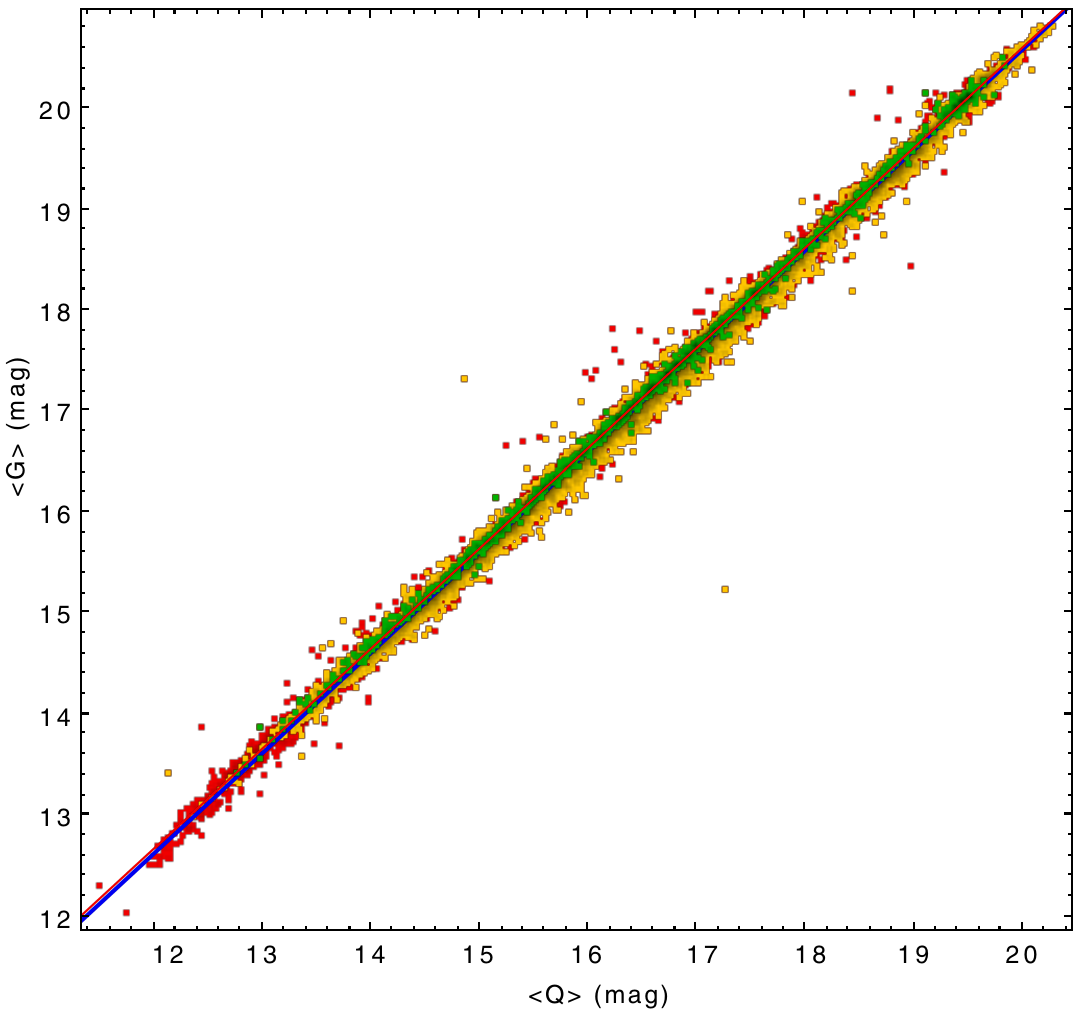}
    \includegraphics[width=0.27\linewidth]{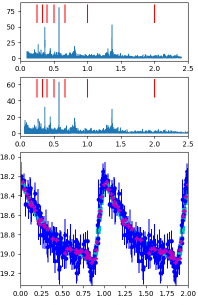}
    \includegraphics[width=0.27\linewidth]{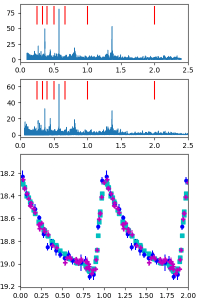}
    \caption{{\it (a - left panel)} Intensity-averaged Gaia G band ($\langle\rm G \rangle$) magnitudes vs. intensity-averaged LSQ Q band ($\langle\rm Q\rangle$) magnitudes for matching objects in the Gaia DR3 and LSQ RRLS catalogs. The {\it red} points show all matches. The {\it yellow} show all matches consistent with being point sources, i.e., that should not have blending issues. The {\it green} are non-blended sources with extinction $A_V < 0.05.$
    {\it (b - middle panel)} Example phase-folded light curves ({\it bottom pane}) and periodograms (AOV, {\it top pane}, and Lomb-Scargle, {\it middle pane}) for an object with good Gaia data and reasonable LSQ data. The {\it violet} points are 
    Gaia data, the {\it dark blue} are LSQ, and the {\it cyan} are the best-fitting template. {\it (c - right panel)} Same as the middle panel except the LSQ data have been binned into 20 phase bins and the median value in those bins is now plotted.
    }
    \label{fig:gaia+lsq}
\end{figure}

To focus on how well we can map Gaia DR3 RRLS photometry into the LSQ system, while avoiding spurious Gaia DR3 objects, we cross-matched the Gaia DR3 RRLS catalog with the LSQ one, which we believe has high purity. A comparison of the intensity-averaged G band magnitude from DR3 vs. the intensity-averaged Q band magnitude from LSQ for 24,147 matched stars is shown in the left panel of Fig. \ref{fig:gaia+lsq}. The LSQ Q band filter is a broad-band (V-centered) filter, which to zeroth order is the sum of the SDSS g and r filters. The Gaia G filter is instead, to the same zeroth order, the sum of the SDSS g,r, and i filters. Since RRLS are blue, with decreasing variability amplitude as one goes to redder wavelengths, one might therefore expect a fairly simple relation to exist between the two filters. This indeed appears to be the case.

Even without quality cuts (see the red points in Fig. \ref{fig:gaia+lsq}), the correlation between averaged Q and G is very good (r=0.998), with a best-fit linear slope of 0.995 and rms dispersion of 0.074 mag. Matching to DELVE DR2 \citep{delvedr2_2022} and including only those objects DELVE considered to be high-confidence point sources (the yellow points, i.e., objects without blending issues), the worst outliers go away and the rms dispersion drops to 0.072. Further requiring that the extinction to a given star be low ($A_V < 0.05$, the green points) decreases the dispersion to 0.055 mag. This dependence on extinction shows that an improved mapping from Gaia to LSQ photometry should include color terms. Even if we ignore such terms, however, the overall agreement between Gaia and LSQ is still quite good considering the systematics involved in deriving intensity-averaged magnitudes from a light curve with a limited number of points that can include bad outliers (estimated to be $\sim 3\%$ for LSQ, and Gaia likely has similar issues).  Also, the LSQ and Gaia data are separated by several years in time, and RRLS pulse profiles can drift on these timescales. Note also that the rms estimate of the scatter about the linear fit is in some ways an overestimate due to non-Gaussian tails. Removing blends but not applying an extinction cut, $\sim 80\%$ of points lie within $\sim 0.05$ mag of the simple linear relation $\langle \rm Q\rangle = \langle \rm G \rangle -0.591,$ comparable to the distance modulus errors discussed above.



Using this relation, we can combine LSQ and Gaia photometry for individual objects, boosting the overall significance of the periodicity detection and improving overall phase coverage of the folded light curve. The middle and right panels of Fig. \ref{fig:gaia+lsq} show an example, for a star with mean magnitude ${\langle \rm V\rangle} \sim 19.6,$ of how Gaia and LSQ light curves can be combined. We chose the Gaia light curve to be a clean example, with good phase coverage, to enable a detailed comparison. The raw LSQ light curve (Fig. \ref{fig:gaia+lsq}b) appears much noisier than the Gaia one, but the LSQ light curve contains many more points ($\sim 200$). Binning in phase (Fig \ref{fig:gaia+lsq}c), we see that the binned LSQ data have comparable errors to the Gaia ones and the LSQ and Gaia data track each other quite well, even at the light curve extremes where the differences caused by LSQ vs. Gaia filter coverage should be greatest. Note that to ensure we stayed on the LSQ photometry system (which this paper is based on), when good (non-blended) LSQ data were available, we added a constant offset to the Gaia data to make the mean magnitude of the Gaia data match the LSQ one before merging the light curves. As seen in Fig. \ref{fig:gaia+lsq}a, the
correction is typically small ($\sim 0.02$ in magnitude for the example star shown in Fig \ref{fig:gaia+lsq}b,c), and even when it is larger, the shifted light curves usually agree very
well (unless the Blazhko effect is significant).

Combining the Gaia light curve data with LSQ data (when available) in this way, we then used our LSQ candidate inspection tool to manually vet all Gaia DR3 candidates fainter than ${\langle \rm G\rangle}=19.34$ ($d_\sun \sim 50{\rm kpc}).$ Significantly more RRc candidate stars were removed by the vetting procedure than RRab candidate stars, which resulted in a ``cleaned" RRab to RRc number ratio of 2.93:1 -- now consistent with LSQ and other halo RRLS studies. The spatial distribution of the vetted DR3 RRLS candidates was also now qualitatively consistent with the LSQ one, with the apparent excess near the Galactic plane gone.

As an end-to-end check of how well we can combine LSQ and Gaia data, we first took the raw Gaia DR3 data alone and matched it to previously published catalogs of type ab and c RRLS in the Draco, Sculptor, and Sextans dwarf galaxies \citep[][respectively] {Bhardwaj2024, martinez2016, vivas2019}, shifted the Gaia $\langle \rm G\rangle$ magnitudes to LSQ $\langle \rm Q\rangle$ ones using the linear relation given above, and then computed the mean distance modulus to those galaxies using the period-luminosity corrections derived above for LSQ. For Draco, 180 type ab/c stars matched, and we obtained a V band distance modulus of $19.583\pm 0.057$ after applying a global extinction correction of $A_V=0.087.$ The distance modulus from \cite{Bhardwaj2024}, based on NIR photometry, is $19.557 \pm 0.026,$ i.e., consistent. For Sculptor, we had 465 type ab/c matches and obtained a distance modulus of
$19.632\pm 0.058$ using a global $A_V=0.051,$ consistent with the literature value of $19.62$ in Table\ref{tableA1}.
For Sextans, we had 79 matches and obtained a distance modulus of $19.711\pm0.121$ using a global $A_V=0.126,$ again consistent with the literature value in Table\ref{tableA1}.

To test how well we can do without relying on prior catalogs to clean the Gaia DR3 sample, we then took our vetted Gaia catalog that included good LSQ light curve data when available and repeated the exercise, this time using a star-by-star V band extinction correction based on \cite{schlafly2011}, as done in the main analysis of this paper. (Note that although 
Gaia light curves may be augmented by LSQ data, this vetted catalog only contains stars found in the original DR3 catalog.) We did not include Draco as it lies outside the LSQ survey area. For Sculptor, the number of matches dropped to 442 and we obtained a mean V band distance modulus of $19.592 \pm 0.127.$ This is consistent with the values in Table \ref{tableA1}, but the rms scatter of the individual RRLS distance-modulus estimates about this mean increased due to the high stellar density in Sculptor affecting LSQ (aperture) photometry more than Gaia photometry. If we remove objects with possible blending issues in LSQ, we have 211 matches and a mean distance modulus of $19.600\pm 0.092,$ closer to the LSQ value.
For Sextans, where blending is not as much of an issue, we have 78 matches and a mean distance modulus of $19.712 \pm 0.100,$ essentially the same as the LSQ result in Table \ref{tableA1}. Overall, the vetting procedure does not significantly reduce the number of matched (previously known) RRLS, and while combining Gaia and LSQ light curves can introduce additional scatter in individual measurements, the mean distance-modulus estimates appear robust and compatible with the LSQ-only results. Hence, it seems justified to add vetted Gaia objects that were missed by LSQ, e.g., due to coverage gaps, to our overall OVOD analysis. The relatively low number of Gaia matches in Sextans \citep[79 out of the 168 type ab/c objects listed in][]{vivas2019} illustrates why it is important to combine LSQ and Gaia catalogs. LSQ has coverage gaps, which Gaia can help fill, but Gaia DR3 by itself has significant completeness/coverage issues, especially at fainter magnitudes.

\bibliography{Outer_VOD}
\bibliographystyle{aasjournalv7}

\end{document}